\documentclass[modern,resetfootnote]{aastex701}

\usepackage{chemformula}
\usepackage{CJK}

\defcitealias{ishiguro16-outbursts}{I16}

\newcommand\mps{m~s$^{-1}$}

\newcommand\kgps{kg~s$^{-1}$}

\newcommand\coo{\mbox{CO$_2$}}

\newcommand\afr{\mbox{$Af\rho$}}

\newcommand\rh{\ensuremath{r_{\mathrm{h}}}}
\newcommand\inv[2][1]{$\textrm{#2}^{-#1}$}

\shorttitle{Outbursts of 7P/Pons--Winnecke}
\shortauthors{Huynh et al.}

\begin{document}
\begin{CJK*}{UTF8}{gbsn}

    \title{A Look at Eight Outbursts of Comet 7P/Pons--Winnecke}

    \correspondingauthor{Michael S. P. Kelley}

    \author[0009-0001-6590-4655]{Ky Huynh}
    \affiliation{Department of Astronomy, University of Maryland, College Park, MD, USA}
    \affiliation{Space Telescope Science Institute, Baltimore, MD, USA}
    \email{kyhuynhh108@gmail.com}

    \author[0000-0002-6702-7676]{Michael S. P. Kelley}
    \affiliation{Department of Astronomy, University of Maryland, College Park, MD, USA}
    \email[show]{msk@astro.umd.edu}

    \author[0000-0002-9413-8785]{Jessica M. Sunshine}
    \affiliation{Department of Astronomy, University of Maryland, College Park, MD, USA}
    \email{jsunshin@umd.edu}

    \author[0000-0002-4838-7676]{Quanzhi Ye (叶泉志)}
    \affiliation{Department of Astronomy, University of Maryland, College Park, MD, USA}
    \affiliation{Center for Space Physics, Boston University, 725 Commonwealth Ave, Boston, MA 02215, USA}
    \email{qye@umd.edu}

    \author[0000-0002-3818-7769]{Tim Lister}
    \affiliation{Las Cumbres Observatory, 6740 Cortona Drive Suite 102, Goleta, CA, USA}
    \email{tlister@lco.global}

    \author[0000-0002-2668-7248]{Dennis Bodewits}
    \affil{Physics Department, Edmund C. Leach Science Center, Auburn University, Auburn, AL, USA}
    \email{dzb0059@auburn.edu}

    \author[0000-0002-0622-2400]{Adam McKay}
    \affil{Department of Physics and Astronomy, Appalachian State University, Boone, NC, USA}
    \email{mckayaj@appstate.edu}

    \author[0000-0003-4365-1455]{Megan E. Schwamb}
    \affiliation{Astrophysics Research Centre, School of Mathematics and Physics, Queen's University Belfast, Belfast BT7 1NN, UK}
    \email{M.Schwamb@qub.ac.uk}

    \author[0000-0002-8658-5534]{Helen Usher}
    \email{helen.usher@open.ac.uk}
    \affiliation{School of Physical Sciences, Open University, Milton Keynes, MK7 6AA, UK}
    \affiliation{School of Physics and Astronomy, Cardiff University, Queens Buildings, The Parade, Cardiff, CF24 3AA, UK}

    \begin{abstract}
        Cometary outbursts may be used as a means to infer the physical processes occurring on cometary nuclei. To that end, we studied eight outbursts of comet 7P/Pons--Winnecke identified between 2021 June 3 and 2021 August 31. The data analyzed consisted of optical images and derived photometry of the comet from the Las Cumbres Observatory network of telescopes. The outburst strengths relative to the ambient coma ranged from --0.2 to --1.1~mag, and the ejecta themselves had apparent brightnesses ranging from 17.4 to 13.3 mag. The morphologies of the ejecta varied, suggesting that the events may have originated from different sources across the nucleus. An order of magnitude estimation of the ejecta masses ranged from 10$^{5}$ - 10$^{6}$ kg, similar to other mini-outbursts of comets. The surface-area normalized outburst rate estimated during this time period is similar to comets 41P/Tuttle--Giacobini--Kres\'ak, 9P/Tempel 1 and 46P/Wirtanen, but 10 times larger than that observed at comet 49P/Arend--Rigaux. However, a comparison to the mini-outburst rate of comet 67P/Churyumov-Gerasimenko reveals significant discrepancies between Rosetta spacecraft results and those from ground-based telescopes.  We also investigate whether or not cometary outbursts from 7P in the 19th century are needed to explain outbursts in meteor shower rates observed in the 20th century.
    \end{abstract}


    \section{Introduction}\label{sec:intro}

    Cometary outbursts are sudden mass-loss events, typically short in duration \citep{hughes90, hughes91, vincent16-fireworks, farnham21-wirtanen}.  Several general processes have been proposed to explain outbursts, including the exothermic annealing of amorphous water ice \citep{sekanina09-crystalization,li11-holmes}, rotationally induced mass wasting \citep{steckloff16-hartley2}, fragmentation \citep{boehnhardt96-sw3,bodewits18-tgk,schleicher19-tgk}, thermal stresses \citep{gronkowski10}, cliff collapse \citep{pajola17}, and explosive release of \coo{} or CO \citep{belton08,moretto17-tempel1}.  The crystallization of amorphous water ice, if present, could be an alternative explanation, as it is thought to release gases more volatile than water, which could drive an outburst \citep{prialnik24-comets3}.  Furthermore, the dust that is ejected can evolve with time \citep{trigo-rodriguez10}, or some events may not even eject any dust \citep{wierzchos20-sw1}.  Identifying the mechanisms that drive cometary outbursts is key to our understanding of the physical processes operating on and within cometary nuclei.

    The smallest events, called mini-outbursts, have been recognized as a potentially common phenomenon among comets, at least since the Deep Impact mission to 9P/Tempel 1 \citep{ahearn05}. Mini-outbursts are events that eject $\lesssim10^6$~kg of dust \citep{belton08,vincent16-fireworks,kelley21-wirtanen}. Comet 9P was regarded as a ``well-behaved'' comet \citep{belton08}, yet under the intense scrutiny of the Deep Impact spacecraft and associated ground-based observing campaign \citep{meech05}, several small outbursts were discovered that likely would have otherwise been missed \citep{lara06,farnham07}.  The Rosetta mission to comet 67P provided our best insight into the causes of the mini-outbursts.  \citet{vincent16-fireworks} measured the source locations of 34 mini-outbursts, and found they were associated with regions of steep topographical structures.  Furthermore, \citet{pajola17} and \citet{agarwal17-outburst} associated separate outbursts with the collapse of specific features.  However, other mechanisms seem to also be at work.  \citet{agarwal17-outburst} additionally studied the dynamics of their event, and concluded that a pressurized subsurface reservoir was needed to explain the amount and acceleration of the dust.  \citet{noonan21-67p-outbursts} characterized two mini-outbursts of 67P, with the first outburst showing indicators of a cliff collapse or similar mass wasting origin and the second more representative of fresh volatiles being exposed via a deepening fracture, followed by two hours of enhanced activity.  Finally, \citet{mueller24-outbursts} identified two types of mini-outbursts with the Rosina mass spectrometer: those driven by water following a cliff collapse, and those driven by explosive \coo{} reservoirs.  They also showed that most of the outbursts of \citet{vincent16-fireworks} were \coo{} driven events.

    Observations of comet 7P/Pons-Winnecke (hereafter, 7P) showed evidence of several outbursts in 2021.  The first observed outburst in 2021 was discovered by \citet{kelley21-atel14486}, from data taken on 2021 March 19 ($\Delta m=-0.38\pm0.12$~mag), 69 days before its perihelion on 2021 May 27 at 02:51 UTC.  A second outburst ($\Delta m=-1.21\pm0.06$~mag) was observed on 2021 June 02 \citep{vanbuitenen21-7p, kelley21-cbet4997}.  These discoveries instigated a follow-up study of the comet by \citet{lister22-look}, as part of the Las Cumbres Observatory Outbursting Objects Key Project (hereafter, LOOK Project).  They found eight additional outbursts occurring between 2021 June 5 and 2021 August 25. Some of the eight outbursts displayed extended features lasting up to a few days before the comet returned to its ambient state. Other outbursts of the comet were identified by T.~Lehmann\footnote{\url{https://groups.io/g/comets-ml/message/29672}}  (2021 May 25, $\sim-0.5$~mag), \citet{sharma21-atel14687} (2021 June 7, $-0.29\pm0.07$~mag), and \citet{kelley22-atel-7p} (2022 November 21, $-1.9\pm0.10$~mag).  In previous orbits, a $\sim-2.5$~mag outburst was discovered by G.~Muler\footnote{\url{https://groups.io/g/comets-ml/message/14245}} in May 2008.

    Regarding the comet itself, 7P is a Jupiter-family comet that was initially discovered by Jean Louis Pons on 1819 June 12, and then rediscovered by Friedrich August Theodor Winnecke on 1858 March 9 \citep{kronk03}. The comet has a 6.3-year orbital period with a perihelion distance of 1.24 au and a minimum orbital intersection distance of $\sim$0.23 au with the Earth.\footnote{Orbital parameters and closest approach data from NASA Jet Propulsion Laboratory's Small-Body Database: \url{https://ssd.jpl.nasa.gov/tools/sbdb_lookup.html}}  7P passed within $\sim$0.2 au of  Earth several times in the late 1800s and early 1900s, and its latest closest approach was on 2021 June 12 when it passed within $\sim$ 0.44 au from Earth, the closest since 1945, making 7P favorable for detailed observations.  Optical images and dust tail models by \citet{mariblanca-escalona25-7p} indicate a dust production rate of approximately 80~\kgps{} near perihelion, and their optical gas production rates are consistent with the typical comet in the carbon-chain taxonomy \citep{ahearn95}.  The comet has a large particle dust trail and an associated meteor shower, the June Bo\"{o}tids \citep{sykes92, tanigawa02-june-bootids}. The nucleus has an estimated radius of 2.6 $\pm$ 0.2 km based on infrared photometry by \citet{fernandez13}, and optical photometry of the nucleus by \citet{snodgrass05} indicate a nuclear axial ratio $a/b\geq1.3 \pm 0.1$ and a rotational period of $\sim$7.9$^{+1.6}_{-1.1}$ hours.  The comet has been considered as a backup flyby target for the European Space Agency's Comet Interceptor spacecraft \citep{schwamb20-ci}.

    This work presents a more detailed analysis of the LOOK Project observations of comet 7P's mini-outbursts.  We examine the photometric and morphological properties of the outburst ejecta, and their frequency.  We also consider if comet 7P's associated meteoroid shower provides evidence for historical outbursts from this comet.

    \section{Observations}
    Our analysis is based on Sloan Digital Sky Survey (SDSS) $r^\prime$ and clear filter images of comet 7P taken by six telescopes that are part of the Las Cumbres Observatory (LCO) global telescope network \citep{brown13-lco}, located at the following six observatories: McDonald Observatory (Texas, USA), South African Astronomical Observatory (Sutherland, South Africa), Teide Observatory (Tenerife, Spain), Haleakala Observatory (Hawaii, USA), Siding Spring Observatory (Coonabarabran, Australia), and Cerro Tololo Interamerican Observatory (Cerro Tololo, Chile).  LOOK Project data were obtained with LCO's 1-m telescopes. Additional data from LCO's 0.4- and 2-m telescopes from the Comet Chasers education and public outreach project were also included in this study. As a result, three different instruments were used: SBIG 6303 cameras with the 0.4-m telescopes (3k$\times$2k, 0\farcs571~\inv{pix}); Sinistro cameras with the 1-m telescopes (4k$\times$4k, 0\farcs389~\inv{pix}); and the $r^\prime$-channel of the 4-channel camera MuSCAT3 on the 2-m Faulkes Telescope North \citep[2k$\times$2k, 0\farcs27~\inv{pix};][]{narita20-muscat3}.  Data are available from the LCO data archive\footnote{\url{https://archive.lco.global/}} under projects KEY2020B-009 (LOOK Project), FTPEPO2014A-004 and FTPEPO2017AB-002 (Comet Chasers).  The LCO archive also has $g$-band observations that were not considered in this work, but were previously presented by \citet{lister22-look}.

\begin{figure}
        \centering
        \includegraphics[width=\textwidth{}]{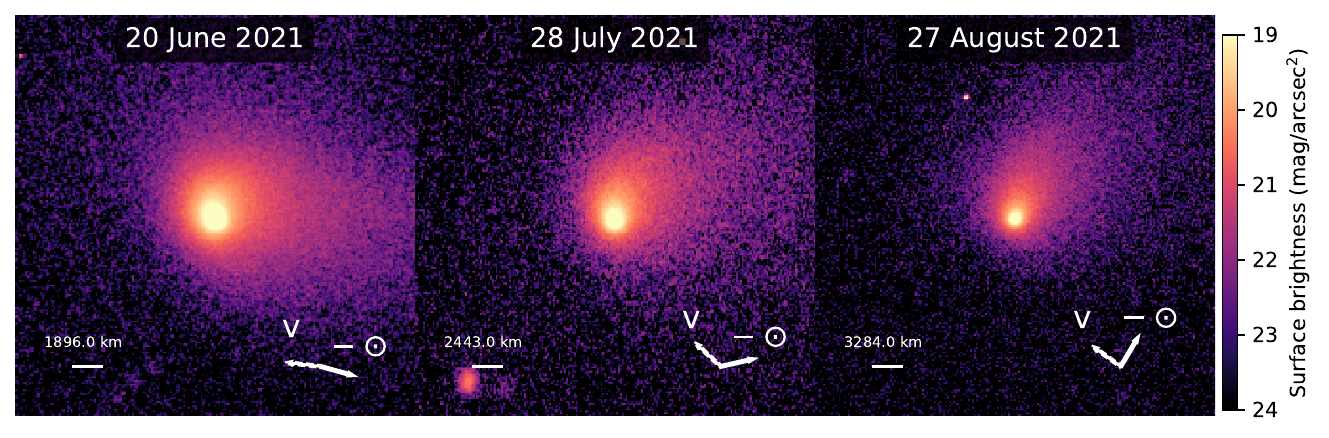}
        \caption{Selected images of comet 7P in an ambient state (dates indicated in each sub-panel). Images are oriented so north is pointing up and east is to the left, and are taken in $r^\prime$ filter. $-\odot$ represents the projected Sun-comet vector and V is the projected heliocentric velocity vector.}
        \label{fig:ambientcomet}
    \end{figure}

    Details on the data acquisition and reduction are presented by \citet{lister22-look}.  In summary, images were taken in observation blocks, which consist of 1--18 exposures (total exposure time per block 60--600~s) at a single telescope.  All observations followed the non-sidereal rates of the comet's ephemeris.  The resulting data were processed through the LCO BANZAI (Beautiful Algorithms to Normalize Zillions of Astronomical Images) pipeline \citep{mccully18-banzai}, which performed standard instrumental calibrations, measured the location and instrumental flux for sources in the image, and generated an astrometric solution. The photometric calibration was calculated with the \texttt{calviacat} software \citep{kelley19-calviacat} by comparing the instrumental magnitudes of background objects to the ATLAS-RefCat2 photometric catalog using the PanSTARRS (PS1) photometric system \citep{tonry18-refcat2}. Color corrections derived from the LOOK Project were used to calibrate the photometry using the comet's measured color: $g-r=0.50\pm0.01$~mag \citep{lister22-look}.  Images were manually inspected and those that have poor calibration stemming from inadequate background removal and/or background star contamination were removed from the analysis. Images were grouped based on filter, observation time, and telescope then averaged together, resulting in 135 images spanning 2021 March 20 to 2021 August 31.   The time between images within an observing block ranged from 8--119~s.  Example images showing comet 7P in its ambient state are presented in Fig.~\ref{fig:ambientcomet} and the observational circumstances of all the images are reported in Table~\ref{tab:observations}.

    \startlongtable
\begin{deluxetable}{lllccccccccc}
    \centerwidetable
    \tablecaption{Observational circumstances and photometry for each image.\label{tab:observations}}
    \tabletypesize{\scriptsize}
    \tablehead{
        \colhead{$N$}
        & \colhead{Obs.\ time}
        & \colhead{Telescope}
        & \colhead{$N_\mathrm{im}$}
        & \colhead{$t_{\mathrm{exp}}$}
        & \colhead{Airmass}
        & \colhead{Seeing}
        & \colhead{\rh{}}
        & \colhead{$\Delta$}
        & \colhead{Phase}
        & \colhead{$r$}
        & \colhead{$\sigma_r$} \\
        & \colhead{(UTC)}
        & 
        & 
        & \colhead{(s)}
        & 
        & \colhead{(")}
        & \colhead{(au)}
        & \colhead{(au)}
        & \colhead{(\degr)}
        & \colhead{(mag)}
        & \colhead{(mag)}
    }
    \colnumbers
    \startdata
    1 & 2021-03-20 11:20 & 1.0 m, McDonald (V39) & 4 & 170.4 & 1.11 & 1.36 & 1.48 & 0.93 & 41.46 & 17.11 & 0.03 \\
    2 & 2021-03-23 01:32 & 1.0 m, Sutherland (K93) & 4 & 170.0 & 1.72 & 1.80 & 1.46 & 0.90 & 42.02 & 17.06 & 0.04 \\
    3 & 2021-03-27 02:32 & 0.4 m, Tenerife (Z21) & 4 & 241.2 & 1.85 & 3.44 & 1.44 & 0.86 & 42.89 & 16.91 & 0.11 \\
    4 & 2021-03-27 14:59 & 0.4 m, Haleakala (T04) & 4 & 239.6 & 1.04 & 1.75 & 1.44 & 0.86 & 43.00 & 16.89 & 0.09 \\
    5 & 2021-03-28 13:54 & 2.0 m, Haleakala (F65) & 6 & 180.0 & 1.14 & 1.13 & 1.43 & 0.84 & 43.21 & 16.98 & 0.09 \\
    6 & 2021-03-28 16:32 & 0.4 m, Siding Spring (Q58) & 1 & 60.2 & 1.84 & 2.37 & 1.43 & 0.84 & 43.24 & 16.81 & 0.10 \\
    7 & 2021-03-29 18:24 & 0.4 m, Siding Spring (Q58) & 1 & 60.2 & 1.36 & 2.43 & 1.42 & 0.83 & 43.48 & 16.56 & 0.19 \\
    8 & 2021-03-30 01:06 & 0.4 m, Sutherland (L09) & 4 & 241.2 & 1.85 & 2.08 & 1.42 & 0.83 & 43.54 & 16.48 & 0.10 \\
    9 & 2021-03-30 14:59 & 0.4 m, Haleakala (T04) & 4 & 239.6 & 1.04 & 1.69 & 1.42 & 0.83 & 43.66 & 16.35 & 0.07 \\
    10 & 2021-03-31 14:59 & 0.4 m, Haleakala (T04) & 4 & 239.6 & 1.04 & 2.48 & 1.41 & 0.82 & 43.89 & 16.35 & 0.08 \\
    11 & 2021-04-01 14:59 & 0.4 m, Haleakala (T04) & 4 & 239.6 & 1.04 & 1.94 & 1.41 & 0.81 & 44.11 & 16.57 & 0.08 \\
    12 & 2021-04-12 17:03 & 2.0 m, Siding Spring (E10) & 6 & 180.0 & 1.53 & 2.25 & 1.35 & 0.71 & 46.58 & 16.21 & 0.11 \\
    13 & 2021-04-23 14:43 & 2.0 m, Haleakala (F65) & 15 & 600.0 & 1.05 & 1.11 & 1.30 & 0.63 & 48.85 & 15.90 & 0.02 \\
    14 & 2021-04-28 14:48 & 2.0 m, Haleakala (F65) & 18 & 540.0 & 1.05 & 0.84 & 1.28 & 0.60 & 49.76 & 15.64 & 0.03 \\
    15 & 2021-06-03 06:05 & 1.0 m, Cerro Tololo (W87) & 6 & 165.0 & 1.62 & 1.94 & 1.24 & 0.45 & 50.68 & 14.54 & 0.04 \\
    16 & 2021-06-03 09:10 & 1.0 m, Cerro Tololo (W86) & 6 & 165.6 & 1.05 & 1.49 & 1.24 & 0.45 & 50.66 & 14.49 & 0.03 \\
    17 & 2021-06-04 00:05 & 1.0 m, Sutherland (K92) & 6 & 165.6 & 1.58 & 2.34 & 1.24 & 0.45 & 50.56 & 14.68 & 0.21 \\
    18 & 2021-06-04 03:01 & 1.0 m, Sutherland (K92) & 6 & 165.6 & 1.06 & 1.84 & 1.24 & 0.45 & 50.54 & 14.65 & 0.05 \\
    19 & 2021-06-04 06:05 & 1.0 m, Cerro Tololo (W86) & 6 & 165.7 & 1.60 & 1.58 & 1.24 & 0.45 & 50.52 & 14.60 & 0.03 \\
    20 & 2021-06-04 09:05 & 1.0 m, Cerro Tololo (W87) & 6 & 165.0 & 1.05 & 1.58 & 1.24 & 0.45 & 50.49 & 14.63 & 0.03 \\
    21 & 2021-06-05 06:05 & 1.0 m, Cerro Tololo (W86) & 6 & 165.6 & 1.59 & 1.83 & 1.24 & 0.45 & 50.32 & 14.46 & 0.02 \\
    22 & 2021-06-05 14:42 & 2.0 m, Haleakala (F65) & 8 & 160.0 & 1.24 & 1.52 & 1.24 & 0.44 & 50.28 & 14.52 & 0.02 \\
    23 & 2021-06-09 06:18 & 1.0 m, Cerro Tololo (W85) & 6 & 165.0 & 1.47 & 2.14 & 1.25 & 0.44 & 49.55 & 14.79 & 0.04 \\
    24 & 2021-06-09 14:21 & 2.0 m, Haleakala (F65) & 4 & 100.0 & 1.31 & 1.45 & 1.25 & 0.44 & 49.48 & 14.56 & 0.03 \\
    25 & 2021-06-11 06:20 & 1.0 m, Cerro Tololo (W87) & 6 & 165.0 & 1.44 & 1.8 & 1.25 & 0.44 & 49.10 & 14.79 & 0.04 \\
    26 & 2021-06-15 05:21 & 1.0 m, Cerro Tololo (W85) & 6 & 165.0 & 1.86 & 2.06 & 1.26 & 0.44 & 48.10 & 14.92 & 0.04 \\
    27 & 2021-06-15 10:03 & 1.0 m, Cerro Tololo (W87) & 6 & 165.0 & 1.02 & 2.00 & 1.26 & 0.44 & 48.04 & 14.50 & 0.05 \\
    28 & 2021-06-16 17:19 & 1.0 m, Siding Spring (Q63) & 6 & 165.0 & 1.09 & 2.15 & 1.26 & 0.44 & 47.69 & 14.90 & 0.03 \\
    29 & 2021-06-17 05:25 & 1.0 m, Cerro Tololo (W85) & 6 & 165.0 & 1.78 & 2.36 & 1.26 & 0.44 & 47.55 & 15.08 & 0.03 \\
    30 & 2021-06-18 04:26 & 1.0 m, Sutherland (K92) & 6 & 165.6 & 1.04 & 1.72 & 1.26 & 0.44 & 47.27 & 14.85 & 0.08 \\
    31 & 2021-06-18 23:08 & 1.0 m, Sutherland (K93) & 6 & 165.0 & 1.84 & 2.06 & 1.27 & 0.44 & 47.04 & 14.75 & 0.04 \\
    32 & 2021-06-19 10:05 & 1.0 m, Cerro Tololo (W85) & 6 & 165.0 & 1.01 & 1.82 & 1.27 & 0.45 & 46.90 & 14.82 & 0.02 \\
    33 & 2021-06-20 05:50 & 1.0 m, Cerro Tololo (W85) & 6 & 165.0 & 1.54 & 2.17 & 1.27 & 0.45 & 46.66 & 14.80 & 0.03 \\
    34 & 2021-06-21 10:35 & 1.0 m, Cerro Tololo (W86) & 6 & 165.6 & 1.03 & 1.42 & 1.27 & 0.45 & 46.29 & 14.81 & 0.02 \\
    35 & 2021-06-22 01:51 & 1.0 m, Sutherland (K91) & 6 & 165.6 & 1.08 & 1.78 & 1.27 & 0.45 & 46.09 & 14.39 & 0.05 \\
    36 & 2021-06-23 00:08 & 1.0 m, Sutherland (K92) & 6 & 165.6 & 1.37 & 1.99 & 1.28 & 0.45 & 45.80 & 14.74 & 0.06 \\
    37 & 2021-06-24 05:52 & 1.0 m, Cerro Tololo (W87) & 2 & 55.0 & 1.48 & 2.11 & 1.28 & 0.45 & 45.40 & 14.88 & 0.05 \\
    38 & 2021-06-24 10:19 & 1.0 m, Cerro Tololo (W86) & 6 & 165.5 & 1.02 & 1.61 & 1.28 & 0.45 & 45.34 & 14.79 & 0.03 \\
    39 & 2021-06-24 23:05 & 1.0 m, Sutherland (K93) & 6 & 165.0 & 1.76 & 2.22 & 1.28 & 0.45 & 45.17 & 14.89 & 0.05 \\
    40 & 2021-06-25 10:03 & 1.0 m, Cerro Tololo (W86) & 6 & 165.6 & 1.01 & 1.35 & 1.29 & 0.46 & 45.01 & 14.77 & 0.06 \\
    41 & 2021-06-26 14:21 & 1.0 m, Siding Spring (Q64) & 6 & 165.6 & 1.85 & 1.64 & 1.29 & 0.46 & 44.63 & 14.78 & 0.05 \\
    42 & 2021-06-27 05:25 & 1.0 m, Cerro Tololo (W87) & 6 & 165.0 & 1.63 & 1.86 & 1.29 & 0.46 & 44.42 & 14.93 & 0.05 \\
    43 & 2021-07-01 05:36 & 1.0 m, Cerro Tololo (W86) & 3 & 82.8 & 1.51 & 1.66 & 1.31 & 0.47 & 43.04 & 14.90 & 0.05 \\
    44 & 2021-07-01 09:53 & 1.0 m, Cerro Tololo (W85) & 4 & 110.0 & 1.01 & 1.76 & 1.31 & 0.47 & 42.97 & 14.99 & 0.04 \\
    45 & 2021-07-02 05:24 & 1.0 m, Cerro Tololo (W86) & 4 & 110.4 & 1.57 & 1.45 & 1.31 & 0.47 & 42.69 & 14.90 & 0.05 \\
    46 & 2021-07-02 14:32 & 1.0 m, Siding Spring (Q63) & 4 & 110.0 & 1.64 & 1.66 & 1.31 & 0.47 & 42.55 & 14.68 & 0.04 \\
    47 & 2021-07-03 10:10 & 1.0 m, Cerro Tololo (W87) & 4 & 110.0 & 1.02 & 1.51 & 1.32 & 0.47 & 42.26 & 14.81 & 0.05 \\
    48 & 2021-07-04 05:08 & 1.0 m, Cerro Tololo (W85) & 4 & 110.0 & 1.66 & 2.07 & 1.32 & 0.48 & 41.98 & 15.02 & 0.05 \\
    49 & 2021-07-04 14:36 & 1.0 m, Siding Spring (Q63) & 4 & 110.0 & 1.58 & 2.70 & 1.32 & 0.48 & 41.84 & 15.03 & 0.04 \\
    50 & 2021-07-05 05:11 & 1.0 m, Cerro Tololo (W87) & 4 & 110.0 & 1.62 & 1.78 & 1.32 & 0.48 & 41.62 & 14.94 & 0.06 \\
    51 & 2021-07-05 15:50 & 1.0 m, Siding Spring (Q64) & 3 & 82.8 & 1.23 & 2.61 & 1.33 & 0.48 & 41.46 & 15.10 & 0.10 \\
    52 & 2021-07-06 10:13 & 1.0 m, Cerro Tololo (W86) & 4 & 110.4 & 1.02 & 2.57 & 1.33 & 0.48 & 41.18 & 14.99 & 0.04 \\
    53 & 2021-07-07 00:25 & 1.0 m, Sutherland (K93) & 4 & 110.0 & 1.21 & 2.92 & 1.33 & 0.48 & 40.97 & 15.07 & 0.04 \\
    54 & 2021-07-07 15:36 & 1.0 m, Siding Spring (Q64) & 4 & 110.4 & 1.26 & 1.60 & 1.34 & 0.49 & 40.74 & 14.97 & 0.06 \\
    55 & 2021-07-07 23:05 & 1.0 m, Sutherland (K91) & 4 & 110.4 & 1.55 & 2.96 & 1.34 & 0.49 & 40.63 & 15.17 & 0.06 \\
    56 & 2021-07-08 10:38 & 1.0 m, Cerro Tololo (W85) & 4 & 110.0 & 1.05 & 1.71 & 1.34 & 0.49 & 40.44 & 15.00 & 0.05 \\
    57 & 2021-07-09 10:39 & 1.0 m, Cerro Tololo (W87) & 4 & 110.0 & 1.06 & 1.77 & 1.34 & 0.49 & 40.08 & 15.02 & 0.05 \\
    58 & 2021-07-10 02:39 & 1.0 m, Sutherland (K93) & 4 & 110.0 & 1.01 & 1.75 & 1.35 & 0.50 & 39.84 & 14.03 & 0.04 \\
    59 & 2021-07-10 10:29 & 1.0 m, Cerro Tololo (W87) & 4 & 110.0 & 1.05 & 1.64 & 1.35 & 0.50 & 39.72 & 14.61 & 0.05 \\
    60 & 2021-07-11 01:10 & 1.0 m, Sutherland (K92) & 4 & 110.4 & 1.09 & 1.96 & 1.35 & 0.50 & 39.50 & 14.88 & 0.05 \\
    61 & 2021-07-11 10:16 & 1.0 m, Cerro Tololo (W87) & 4 & 110.0 & 1.04 & 1.52 & 1.35 & 0.50 & 39.35 & 14.94 & 0.04 \\
    62 & 2021-07-11 23:01 & 1.0 m, Sutherland (K92) & 4 & 110.4 & 1.51 & 2.39 & 1.36 & 0.50 & 39.16 & 15.07 & 0.05 \\
    63 & 2021-07-13 04:49 & 1.0 m, Cerro Tololo (W86) & 4 & 110.4 & 1.64 & 2.07 & 1.36 & 0.51 & 38.71 & 15.06 & 0.03 \\
    64 & 2021-07-13 10:06 & 1.0 m, Cerro Tololo (W85) & 4 & 110.0 & 1.04 & 2.86 & 1.36 & 0.51 & 38.62 & 15.18 & 0.04 \\
    65 & 2021-07-14 04:47 & 1.0 m, Cerro Tololo (W85) & 4 & 110.0 & 1.64 & 2.42 & 1.37 & 0.51 & 38.34 & 15.19 & 0.04 \\
    66 & 2021-07-14 09:53 & 1.0 m, Cerro Tololo (W85) & 5 & 134.4 & 1.03 & 1.77 & 1.37 & 0.51 & 38.26 & 15.13 & 0.04 \\
    67 & 2021-07-14 22:37 & 1.0 m, Sutherland (K91) & 4 & 110.4 & 1.61 & 1.90 & 1.37 & 0.51 & 38.07 & 14.97 & 0.04 \\
    68 & 2021-07-15 09:53 & 1.0 m, Cerro Tololo (W85) & 4 & 110.0 & 1.03 & 1.94 & 1.37 & 0.52 & 37.90 & 15.03 & 0.04 \\
    69 & 2021-07-15 22:40 & 1.0 m, Sutherland (K93) & 4 & 110.0 & 1.57 & 1.86 & 1.38 & 0.52 & 37.71 & 15.02 & 0.05 \\
    70 & 2021-07-16 09:53 & 1.0 m, Cerro Tololo (W85) & 4 & 110.0 & 1.04 & 1.91 & 1.38 & 0.52 & 37.53 & 15.08 & 0.05 \\
    71 & 2021-07-16 23:01 & 1.0 m, Sutherland (K91) & 4 & 110.4 & 1.44 & 2.07 & 1.38 & 0.52 & 37.34 & 15.10 & 0.05 \\
    72 & 2021-07-18 02:16 & 1.0 m, Sutherland (K91) & 4 & 110.4 & 1.02 & 1.50 & 1.39 & 0.53 & 36.92 & 15.08 & 0.05 \\
    73 & 2021-07-18 09:53 & 1.0 m, Cerro Tololo (W85) & 4 & 110.0 & 1.04 & 1.77 & 1.39 & 0.53 & 36.81 & 15.14 & 0.06 \\
    74 & 2021-07-18 22:17 & 1.0 m, Sutherland (K93) & 4 & 110.0 & 1.67 & 2.33 & 1.39 & 0.53 & 36.63 & 14.02 & 0.04 \\
    75 & 2021-07-19 09:53 & 1.0 m, Cerro Tololo (W85) & 4 & 110.0 & 1.05 & 1.82 & 1.39 & 0.53 & 36.45 & 14.89 & 0.04 \\
    76 & 2021-07-19 22:30 & 1.0 m, Sutherland (K91) & 4 & 110.4 & 1.56 & 1.64 & 1.40 & 0.54 & 36.26 & 15.00 & 0.05 \\
    77 & 2021-07-20 09:53 & 1.0 m, Cerro Tololo (W85) & 4 & 110.0 & 1.05 & 1.87 & 1.40 & 0.54 & 36.09 & 15.12 & 0.04 \\
    78 & 2021-07-23 04:17 & 1.0 m, Cerro Tololo (W85) & 4 & 110.0 & 1.66 & 1.57 & 1.42 & 0.55 & 35.12 & 15.18 & 0.06 \\
    79 & 2021-07-23 10:25 & 1.0 m, Cerro Tololo (W86) & 4 & 110.4 & 1.10 & 1.33 & 1.42 & 0.55 & 35.02 & 15.14 & 0.05 \\
    80 & 2021-07-25 04:29 & 1.0 m, Cerro Tololo (W85) & 4 & 110.0 & 1.54 & 2.54 & 1.43 & 0.56 & 34.42 & 15.32 & 0.05 \\
    81 & 2021-07-25 14:02 & 1.0 m, Siding Spring (Q63) & 4 & 110.0 & 1.46 & 1.63 & 1.43 & 0.56 & 34.28 & 15.15 & 0.05 \\
    82 & 2021-07-26 10:25 & 1.0 m, Cerro Tololo (W85) & 4 & 110.0 & 1.12 & 1.97 & 1.43 & 0.57 & 33.99 & 15.31 & 0.06 \\
    83 & 2021-07-26 22:25 & 1.0 m, Sutherland (K93) & 4 & 110.0 & 1.46 & 3.18 & 1.44 & 0.57 & 33.83 & 15.30 & 0.04 \\
    84 & 2021-07-27 10:27 & 1.0 m, Cerro Tololo (W87) & 4 & 110.0 & 1.13 & 1.89 & 1.44 & 0.57 & 33.65 & 15.19 & 0.05 \\
    85 & 2021-07-27 22:25 & 1.0 m, Sutherland (K91) & 4 & 110.4 & 1.45 & 1.92 & 1.44 & 0.57 & 33.49 & 15.22 & 0.05 \\
    86 & 2021-07-28 10:25 & 1.0 m, Cerro Tololo (W85) & 4 & 110.0 & 1.13 & 1.92 & 1.45 & 0.58 & 33.32 & 15.24 & 0.04 \\
    87 & 2021-07-29 04:25 & 1.0 m, Sutherland (K92) & 4 & 110.5 & 1.14 & 2.16 & 1.45 & 0.58 & 33.07 & 15.30 & 0.04 \\
    88 & 2021-07-29 10:25 & 1.0 m, Cerro Tololo (W85) & 4 & 110.0 & 1.14 & 1.74 & 1.45 & 0.58 & 32.99 & 15.21 & 0.04 \\
    89 & 2021-07-29 22:51 & 1.0 m, Sutherland (K93) & 4 & 110.0 & 1.32 & 2.98 & 1.46 & 0.58 & 32.83 & 15.29 & 0.04 \\
    90 & 2021-07-30 23:52 & 1.0 m, Sutherland (K93) & 4 & 110.0 & 1.15 & 1.75 & 1.46 & 0.59 & 32.49 & 15.31 & 0.04 \\
    91 & 2021-07-31 10:25 & 1.0 m, Cerro Tololo (W87) & 4 & 110.0 & 1.16 & 1.46 & 1.46 & 0.59 & 32.34 & 15.20 & 0.06 \\
    92 & 2021-08-03 01:41 & 1.0 m, Sutherland (K91) & 4 & 168.4 & 1.04 & 1.74 & 1.48 & 0.61 & 31.52 & 15.25 & 0.05 \\
    93 & 2021-08-04 03:39 & 1.0 m, Sutherland (K91) & 4 & 168.4 & 1.11 & 1.52 & 1.49 & 0.61 & 31.19 & 15.23 & 0.05 \\
    94 & 2021-08-05 02:15 & 1.0 m, Sutherland (K92) & 4 & 168.4 & 1.04 & 1.94 & 1.49 & 0.62 & 30.91 & 15.31 & 0.05 \\
    95 & 2021-08-06 00:24 & 1.0 m, Sutherland (K91) & 4 & 168.4 & 1.08 & 3.26 & 1.50 & 0.62 & 30.64 & 15.62 & 0.18 \\
    96 & 2021-08-07 00:55 & 1.0 m, Sutherland (K92) & 4 & 168.4 & 1.05 & 1.92 & 1.51 & 0.63 & 30.35 & 15.35 & 0.04 \\
    97 & 2021-08-07 15:52 & 1.0 m, Siding Spring (Q63) & 3 & 126.0 & 1.08 & 6.83 & 1.51 & 0.63 & 30.18 & 15.61 & 0.09 \\
    98 & 2021-08-08 00:24 & 1.0 m, Sutherland (K91) & 4 & 168.4 & 1.07 & 1.77 & 1.51 & 0.64 & 30.08 & 15.37 & 0.05 \\
    99 & 2021-08-08 17:45 & 1.0 m, Siding Spring (Q63) & 4 & 168.0 & 1.06 & 2.52 & 1.52 & 0.64 & 29.87 & 15.41 & 0.07 \\
    100 & 2021-08-09 06:32 & 1.0 m, Cerro Tololo (W87) & 4 & 168.0 & 1.09 & 1.67 & 1.52 & 0.64 & 29.73 & 15.33 & 0.04 \\
    101 & 2021-08-10 00:24 & 1.0 m, Sutherland (K91) & 4 & 168.4 & 1.07 & 1.47 & 1.53 & 0.65 & 29.53 & 15.38 & 0.05 \\
    102 & 2021-08-10 12:41 & 1.0 m, Siding Spring (Q63) & 4 & 168.0 & 1.55 & 2.17 & 1.53 & 0.65 & 29.40 & 15.47 & 0.05 \\
    103 & 2021-08-11 00:46 & 1.0 m, Sutherland (K91) & 4 & 168.4 & 1.06 & 1.72 & 1.53 & 0.66 & 29.26 & 15.43 & 0.05 \\
    104 & 2021-08-11 17:08 & 1.0 m, Siding Spring (Q64) & 3 & 126.3 & 1.06 & 2.29 & 1.54 & 0.66 & 29.09 & 15.45 & 0.08 \\
    105 & 2021-08-12 03:10 & 1.0 m, Cerro Tololo (W86) & 4 & 168.4 & 1.59 & 1.64 & 1.54 & 0.66 & 28.98 & 15.43 & 0.05 \\
    106 & 2021-08-13 04:35 & 1.0 m, Cerro Tololo (W87) & 4 & 168.0 & 1.25 & 2.00 & 1.55 & 0.67 & 28.72 & 15.46 & 0.04 \\
    107 & 2021-08-13 17:44 & 1.0 m, Siding Spring (Q63) & 4 & 168.0 & 1.08 & 1.82 & 1.55 & 0.67 & 28.58 & 15.50 & 0.04 \\
    108 & 2021-08-14 06:03 & 1.0 m, Cerro Tololo (W87) & 4 & 168.0 & 1.10 & 1.72 & 1.56 & 0.68 & 28.46 & 15.50 & 0.04 \\
    109 & 2021-08-14 13:00 & 1.0 m, Siding Spring (Q63) & 4 & 168.0 & 1.39 & 1.37 & 1.56 & 0.68 & 28.40 & 15.48 & 0.03 \\
    110 & 2021-08-15 15:09 & 1.0 m, Siding Spring (Q64) & 4 & 168.4 & 1.10 & 2.06 & 1.56 & 0.68 & 28.14 & 15.57 & 0.04 \\
    111 & 2021-08-16 02:53 & 1.0 m, Cerro Tololo (W86) & 4 & 168.4 & 1.58 & 1.63 & 1.57 & 0.69 & 28.03 & 15.54 & 0.05 \\
    112 & 2021-08-16 12:43 & 1.0 m, Siding Spring (Q63) & 4 & 168.0 & 1.42 & 1.84 & 1.57 & 0.69 & 27.94 & 15.56 & 0.05 \\
    113 & 2021-08-17 00:24 & 1.0 m, Sutherland (K91) & 4 & 168.4 & 1.06 & 1.55 & 1.57 & 0.70 & 27.83 & 15.55 & 0.06 \\
    114 & 2021-08-17 12:24 & 1.0 m, Siding Spring (Q63) & 4 & 168.0 & 1.47 & 2.13 & 1.58 & 0.70 & 27.73 & 15.66 & 0.05 \\
    115 & 2021-08-18 02:32 & 1.0 m, Cerro Tololo (W86) & 4 & 168.4 & 1.64 & 1.74 & 1.58 & 0.70 & 27.60 & 15.60 & 0.05 \\
    116 & 2021-08-19 12:30 & 1.0 m, Siding Spring (Q64) & 4 & 168.4 & 1.41 & 1.58 & 1.59 & 0.71 & 27.31 & 15.58 & 0.06 \\
    117 & 2021-08-21 02:17 & 1.0 m, Cerro Tololo (W85) & 4 & 168.0 & 1.64 & 1.81 & 1.60 & 0.72 & 27.01 & 15.69 & 0.07 \\
    118 & 2021-08-21 12:45 & 1.0 m, Siding Spring (Q64) & 4 & 168.4 & 1.33 & 1.52 & 1.60 & 0.73 & 26.93 & 15.62 & 0.06 \\
    119 & 2021-08-22 05:01 & 1.0 m, Cerro Tololo (W86) & 4 & 168.4 & 1.13 & 1.50 & 1.61 & 0.73 & 26.80 & 15.58 & 0.06 \\
    120 & 2021-08-23 05:52 & 1.0 m, Cerro Tololo (W86) & 4 & 168.5 & 1.08 & 1.52 & 1.62 & 0.74 & 26.62 & 15.61 & 0.06 \\
    121 & 2021-08-24 00:27 & 1.0 m, Sutherland (K92) & 4 & 168.3 & 1.06 & 1.91 & 1.62 & 0.75 & 26.49 & 15.76 & 0.05 \\
    122 & 2021-08-24 20:16 & 1.0 m, Sutherland (K93) & 4 & 168.0 & 1.48 & 2.01 & 1.63 & 0.75 & 26.36 & 15.79 & 0.06 \\
    123 & 2021-08-25 00:55 & 1.0 m, Sutherland (K92) & 4 & 168.4 & 1.07 & 1.71 & 1.63 & 0.75 & 26.33 & 15.76 & 0.07 \\
    124 & 2021-08-25 15:24 & 1.0 m, Siding Spring (Q63) & 4 & 168.0 & 1.07 & 2.08 & 1.63 & 0.76 & 26.24 & 15.60 & 0.06 \\
    125 & 2021-08-26 02:23 & 1.0 m, Sutherland (K92) & 4 & 168.4 & 1.16 & 2.37 & 1.64 & 0.76 & 26.17 & 15.75 & 0.06 \\
    126 & 2021-08-27 01:47 & 1.0 m, Cerro Tololo (W85) & 4 & 168.0 & 1.64 & 1.74 & 1.64 & 0.77 & 26.03 & 15.82 & 0.04 \\
    127 & 2021-08-27 13:25 & 1.0 m, Siding Spring (Q63) & 4 & 168.0 & 1.16 & 1.40 & 1.65 & 0.77 & 25.96 & 15.82 & 0.04 \\
    128 & 2021-08-28 01:47 & 1.0 m, Cerro Tololo (W85) & 4 & 168.0 & 1.61 & 1.76 & 1.65 & 0.78 & 25.89 & 15.91 & 0.04 \\
    129 & 2021-08-28 18:36 & 1.0 m, Siding Spring (Q64) & 4 & 148.5 & 1.31 & 1.56 & 1.66 & 0.78 & 25.80 & 15.91 & 0.05 \\
    130 & 2021-08-28 19:56 & 1.0 m, Sutherland (K93) & 4 & 168.0 & 1.47 & 2.45 & 1.66 & 0.78 & 25.79 & 15.89 & 0.10 \\
    131 & 2021-08-29 00:29 & 1.0 m, Sutherland (K93) & 4 & 168.0 & 1.07 & 2.57 & 1.66 & 0.79 & 25.77 & 15.91 & 0.09 \\
    132 & 2021-08-29 15:01 & 1.0 m, Siding Spring (Q63) & 4 & 168.0 & 1.07 & 1.65 & 1.66 & 0.79 & 25.69 & 15.96 & 0.05 \\
    133 & 2021-08-30 02:55 & 1.0 m, Cerro Tololo (W86) & 4 & 168.4 & 1.30 & 1.56 & 1.67 & 0.80 & 25.64 & 15.98 & 0.05 \\
    134 & 2021-08-30 14:21 & 1.0 m, Siding Spring (Q63) & 4 & 168.0 & 1.09 & 1.33 & 1.67 & 0.80 & 25.58 & 15.99 & 0.06 \\
    135 & 2021-08-31 04:22 & 1.0 m, Cerro Tololo (W85) & 4 & 168.0 & 1.13 & 1.90 & 1.67 & 0.80 & 25.51 & 16.06 & 0.05 \\
    \enddata
    \tablecomments{All but one image were taken in the $r^\prime$ filter of their respective instruments ($N=22$ was unfiltered).  Columns: (1) row number, (2) mid-time of observation, (3) telescope, site, and MPC code, (4) number of images combined, (5) exposure time, (6) airmass, (7) estimated seeing (FWHM of point sources) conditions, (8) heliocentric distance, (9) comet-observer distance, (10) Sun-comet-observer (phase) angle, (11) $r$-band apparent magnitude (PS1 system), (12) uncertainty in $r$-band apparent magnitude.}
\end{deluxetable}

    \section{Results and Analysis}
    \subsection{Photometry and outburst identification}
    Photometry was measured from the stacked images using an aperture radius of 3\farcs9 (1200--2600~km) centered on the inner coma.  The choice of aperture size affects the discoverability of outbursts.  Smaller apertures reduce the amount of ambient coma and thereby enhance the contrast of the outbursts in the lightcurve, provided that the outburst material is observed near the nucleus.  However, the sky conditions (seeing), which vary from observation to observation, will add an additional noise source for apertures smaller than or similar to the full width-half-maximum (FWHM) of point sources.  The seeing is better than 2\farcs5 in 90\% of the images, and therefore the effects on our our 3\farcs9 aperture radius photometry is limited.  The results are converted to absolute magnitude, $H$, using the equation:
    \begin{equation}
        H=m - 5 \log_{10}(\rh\,\Delta) + 2.5 \log_{10} \Phi(\phi)
    \end{equation}
    where $\Phi$ is the phase function of the coma evaluated at phase (Sun-comet-observer) angle $\phi$.  For the phase function, we assume the brightness is dominated by dust and use the Schleicher-Marcus phase function normalized at 0\degr{} \citep{schleicher98,marcus07b} derived from 1P/Halley over the observed geometry for 7P ($\phi=26-51\degr$). Error sources include photometric calibrations (based on the standard deviation of the background sources about the best fit solution), background noise, and telescope sensitivity, all of which contribute to the average uncertainty of $\sim$0.05 mag on the photometric data.

\begin{figure}[t]
        \includegraphics[width=\textwidth]{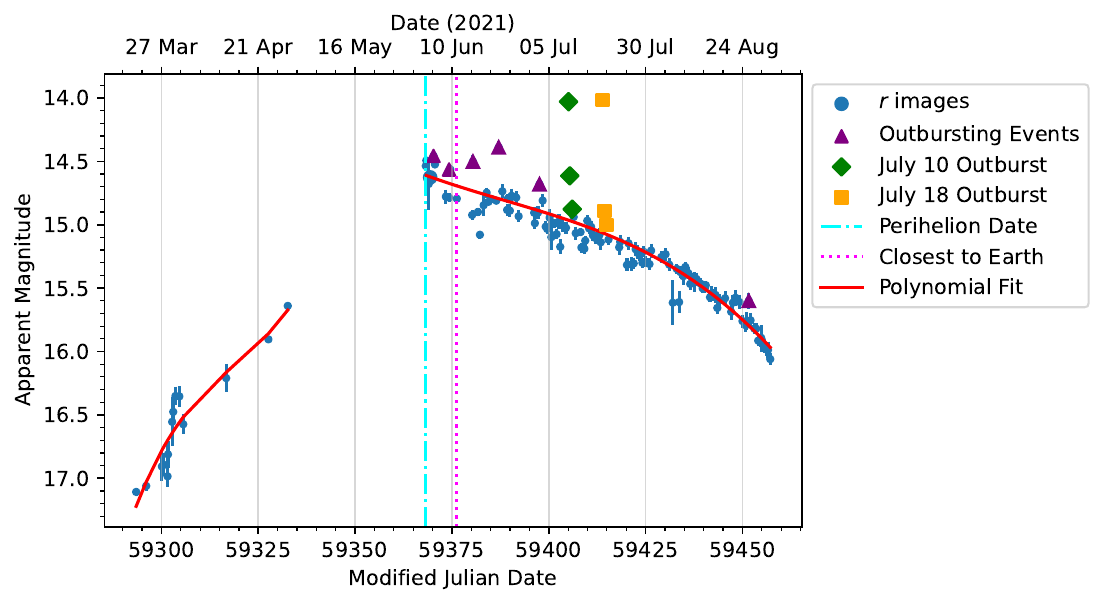}
        \caption{Light curve of comet 7P/Pons--Winnecke based on $r$-band photometry (blue circles). Purple triangles indicate an identified post-perihelion outburst, except for the outbursts of July 10 and 18, which are marked with green diamonds and yellow squares, respectively. The cyan dash-dotted and pink dotted lines indicate the time of perihelion and closest approach to Earth, respectively. Red solid lines are third-order polynomial fits to the pre- and post-perihelion photometry used to identify candidate outbursts.}
        \label{fig:lightcurve}
    \end{figure}

    Figure \ref{fig:lightcurve} presents the photometry of 7P versus time.  The pre- and post-perihelion data were fit with a third-order polynomial to define the general light curve trends. Outliers that deviated from the polynomial trend by 3$\sigma$ were considered as candidate outbursts. Images for each candidate were visually inspected to determine if the photometry was affected by artifacts or background objects. As a further test, the most recent image of the coma in quiescence was subtracted from the candidate to determine if any new ejecta could account for the change in brightness.  For fainter events, we additionally compared the images to Digitized Sky Survey data from the Mikulski Archive for Space Telescopes to verify that the candidates were not affected by background sources.  Candidates with detectable outburst ejecta were classified as confirmed outbursts.

    Using this method, eight outbursts were identified between 2021 June 5 to August 25. The timing of the outbursts, the change in absolute magnitude, lower limit of the expansion speed, dust mass, observational circumstances, and the apparent magnitude of the ejecta are listed in Table~\ref{tab:outbursts}. These are the same outbursts identified by \citet{lister22-look}.  The strongest outbursts in the data set occurred on July 10 and July 18, both increasing the brightness by approximately --1 mag.  The ejecta from each of these events are apparent in our photometric aperture for 44 and 36 hours, respectively.  The remaining weaker outbursts were only apparent for $\sim$1 day before the light curve returned to its ambient state.

    The time-averaged outburst rate is 0.090~\inv{day} for the 89 day high-cadence period.  Adding the May 25, June 2, and June 7 outbursts reported elsewhere (Section~\ref{sec:intro}), the rate increases slightly to 0.10~\inv{day} between 2021 May 15 to August 31.  The elapsed time between outbursts varied.  For the first seven events in our data set, the time between events ranged from 4.4 to 10.5 days (mean 7.3 days), and the eighth outburst followed 37.7 days after the 7th event.  However, the long delay before the 8th outburst may not be physically significant.  A Kolmogorov-Smirnov test indicates the elapsed time between outbursts is statistically indistinguishable from a distribution that is uniform in time ($p$-value = 0.21).

    Given the fast light curve evolution of the outbursts, and the five multi-day gaps in temporal coverage in our light curve, some events could have been missed.  Based on an average outburst rate of 0.10~\inv{day}, and the 15 cumulative days of no data during five multi-day gaps in the data set, we calculate that 1.5 outbursts may have been missed by the LCO data set.  Assuming a Poisson distribution, there is a 2\% chance that more than 4 outbursts were missed.  Therefore, we take the number of outbursts during the period of 2021 Jun 3 to Aug 31 to be 8 to 12, corresponding to a rate of 0.09 to 0.13~\inv{day} (98\% confidence interval).

    We produced images of the ejecta by subtracting a preceding image of the comet in quiescence from each image of the comet in outburst. The time between the outburst image and the preceding image ranged from 4.5 to 21.0 hours, which we define as the maximum age of the outburst. The apparent brightness of the outburst material was measured from the difference image, and the result is converted into an absolute magnitude.

\begin{figure}
        \includegraphics[width=\textwidth]{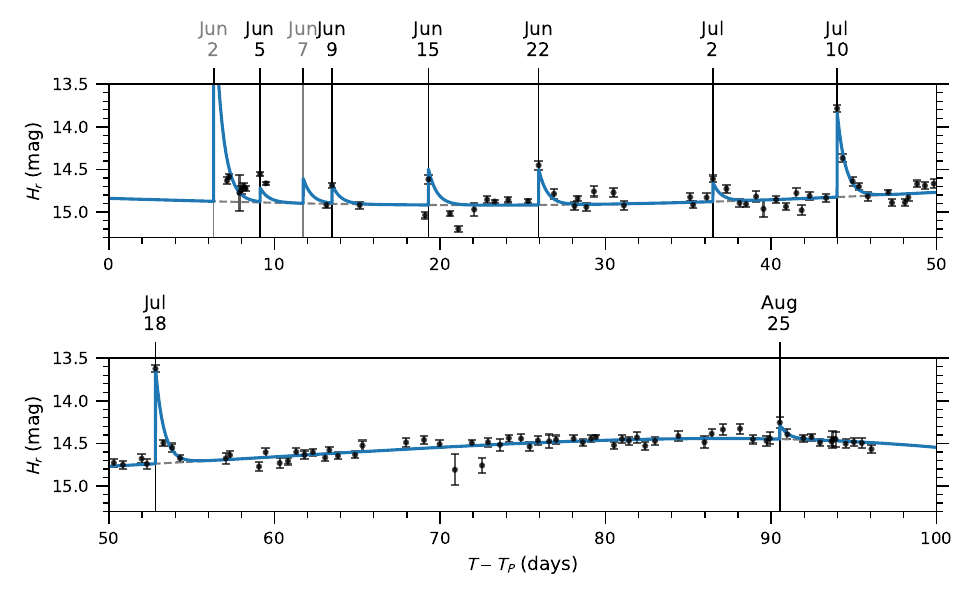}
        \caption{Absolute $r$-band magnitude of comet 7P/Pons--Winnecke versus time from perihelion.  A polynomial fit to the quiescent activity is shown (dashed line).  Outbursts are marked with vertical lines, including the June 2 \citep{vanbuitenen21-7p} and June 7 \citep{sharma21-atel14687} outbursts, which are labeled in gray.  Illustrative models for each outburst's evolution are shown as exponential functions (solid line).}
        \label{fig:outburst-lightcurves}
    \end{figure}

    We compared the shapes of the post-outburst light curves to an exponential function of the form:
    \begin{equation}
        H_r = H_0 + \Delta H \exp{(-t/\tau)}
        \label{eqn:expfunc}
    \end{equation}
    where $H_r$ is the $r$-band absolute magnitude of the comet in the 3\farcs9 radius aperture, $\Delta H$ is the strength of the outburst from Table~\ref{tab:outbursts}, $t$ is the time since the outburst discovery, and $\tau$ is a timescale that accounts for the rate of loss of material from the photometric aperture.  Based on a manual comparison to the June 22, July 2, 10, and 18 events, $\tau=0.5$~days best represented the data. Fig.~\ref{fig:outburst-lightcurves} shows the light curve with a fourth order polynomial fitted to the data.  The polynomial order was the minimum required to achieve a satisfactory fit to the data.  Superimposed on the polynomial are outburst light curves using our exponential function.  The times of the June 2 and June 7 outbursts of \citet{vanbuitenen21-7p} and \citet{sharma21-atel14687} are also shown.  There are some discrepancies with the early data.  These are caused by a difficulty establishing the light curve trend due to the strong June 2 outburst and the low data cadence before June 15.

    \begin{deluxetable}{lcCCCCCCCC}
        \tablecaption{Outburst timing, brightness, expansion speed, and dust mass. \label{tab:outbursts}}
        \colnumbers
        \tablehead{
            \colhead{Time}
            & \colhead{$t_\mathrm{max}$}
            & \colhead{$\Delta H$}
            & \colhead{$\sigma_{\Delta H}$}
            & \colhead{$\rho_\mathrm{max}$}
            & \colhead{$m_\mathrm{ejecta}$}
            & \colhead{$\sigma_m$}
            & \colhead{Mass}
            & \colhead{Mass (M25)}
            & \colhead{Min. Expansion Speed} \\
            \colhead{(UTC)}
            & \colhead{(hr)}
            & \colhead{(mag)}
            & \colhead{(mag)}
            & \colhead{(\arcsec)}
            & \colhead{(mag)}
            & \colhead{(mag)}
            & \colhead{(kg)}
            & \colhead{(kg)}
            & \colhead{(m/s)}
        }
        \tablecolumns{8}
        \startdata
        2021 Jun  5\tablenotemark{a} & 21.0 & -0.16 & 0.06 &  3.96 & 16.20 & 0.04 & 1.2 \times 10^5 & 2.6 \times 10^6 & 17 \\
        2021 Jun  9 & 8.05 & -0.21 & 0.07 &  2.66 & 15.47 & 0.04 & 2.2 \times 10^5 & 4.9 \times 10^6 & 29 \\
        2021 Jun 15 & 4.70 & -0.42 & 0.11 &  3.90 & 15.64 & 0.06 & 2.0 \times 10^5 & 4.3 \times 10^6 & 74 \\
        2021 Jun 22 & 15.3 & -0.42 & 0.10 &  6.46 & 15.15 & 0.07 & 3.1 \times 10^5 & 7.0 \times 10^6 & 38 \\
        2021 Jul  2 & 9.13 & -0.22 & 0.10 &  4.19 & 16.22 & 0.12 & 1.3 \times 10^5 & 3.0 \times 10^6 & 44 \\
        2021 Jul 10 & 16.0 & -0.99 & 0.10 & 10.83 & 13.26 & 0.04 & 2.1 \times 10^6 & 5.3 \times 10^7 & 67 \\
        2021 Jul 18 & 12.4 & -1.12 & 0.11 &  7.77 & 13.90 & 0.04 & 1.5 \times 10^6 & 3.4 \times 10^7 & 67 \\
        2021 Aug 25 & 14.5 & -0.16 & 0.12 &  4.14 & 17.43 & 0.20 & 1.4 \times 10^5 & 3.2 \times 10^6 & 44
        \enddata
        \tablecomments{Columns: (1) time of the initial outburst detection, (2) maximum age of the outburst at the time of discovery, (3) change in absolute brightness in our nominal 3\farcs9 radius aperture based on the absolute magnitude, (4) uncertainty on $\Delta H$, (5) radius of a circle that circumscribes the area used to measure the total ejecta brightness, (6) apparent magnitude of the ejecta, (7) uncertainty on $m_\mathrm{ejecta}$, (8) mass of the ejecta, (9) mass of the ejecta given the dust parameters of \citet{mariblanca-escalona25-7p}}, (10) lower limit of the expansion speed.
        \tablenotetext{a}{The amount of material for the June 5 outburst is likely underestimated due to a difficultly establishing the brightness before the event.}
    \end{deluxetable}

    \subsection{Ejecta morphology}

\begin{figure}[t]
        \centering
        \includegraphics[width=\textwidth]{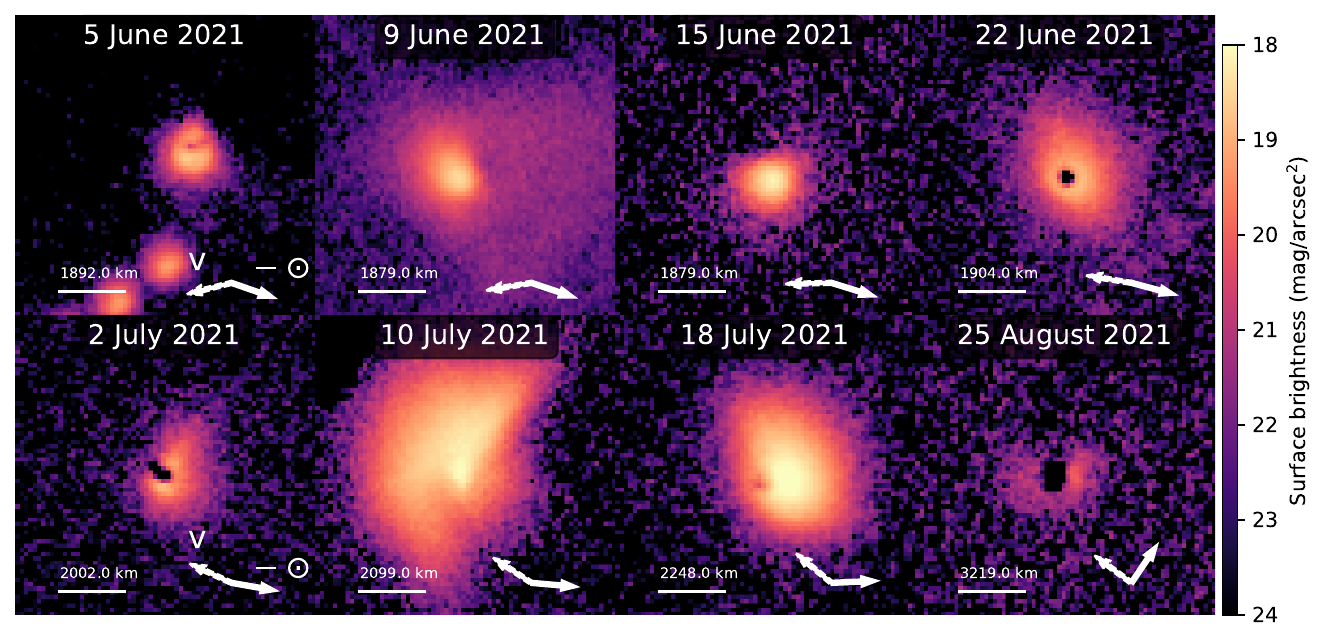}
        \caption{Images of the outburst ejecta generated from the difference between an outburst image and the image preceding it. Images are oriented so that north is up and east is to the left. All displayed images are taken in $r^\prime$ filter. $-\odot$ represents the anti-solar direction and V is the heliocentric velocity direction.  Additional compact features in the lower half of the June 5 image are due to background sources.}
        \label{fig:outburst-images}
    \end{figure}

\begin{figure}
        \centering
        \includegraphics[scale=0.40]{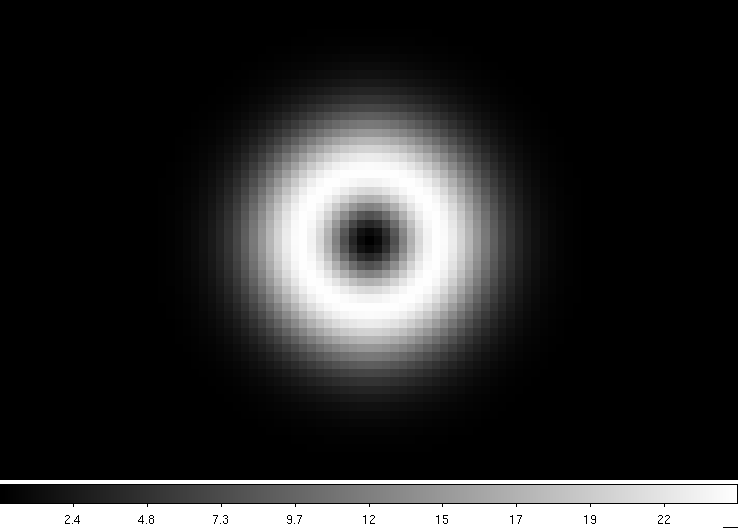}
        \caption{Difference of two 2D Gaussian distributions, with $\sigma$=5 and 7~pix. The artifacts seen in the June 5, June 22, July 2, and August 25 outbursts are replicated, indicating that seeing effects are the likely cause.}
        \label{fig:seeing}
    \end{figure}

    Figure \ref{fig:outburst-images} shows our derived images of the outburst ejecta.  The resulting June 5, June 22, July 2, and August 25 ejecta images have a processing artifact, apparent as a dark region centered on the nucleus.  This artifact is a result of the seeing difference between the two images.  Atmospheric conditions as well as which telescope is used varied between observational epochs.  Images that contain the artifact were found to have a seeing difference of $\geq$ 0\farcs3 between the outburst image and its preceding image. Images that did not have the artifact have a seeing difference $\leq$ 0\farcs1.  We identified pairs of images of the ambient coma with similar seeing conditions and examined the difference.  The artifact did not appear in these test cases.  The effect of a change in seeing is demonstrated in Fig.~\ref{fig:seeing}, where we plot the difference of two Gaussian distributions with $\sigma$ = 7~pix and $\sigma$ = 5~pix.  The morphology of the difference (a dark region surrounded by a ring) is similar to the dark artifacts in our ejecta images.  Despite the appearance, the artifact does not play a significant role in the photometry of the ejecta as seeing variations should preserve the total brightness.

    The morphology of each event varies, and four events are bright enough for detailed analysis: June 9, June 22, July 10, and July 18.  The ejecta distribution tends to be anisotropic about the nucleus. However, there is no preferred direction overall, e.g., the June 9 and July 10 outbursts are directed more towards the Sun, but the June 22 and July 18 outbursts are more directed towards the tail.

\begin{figure}[ht]
        \centering
        \includegraphics[scale=.50]{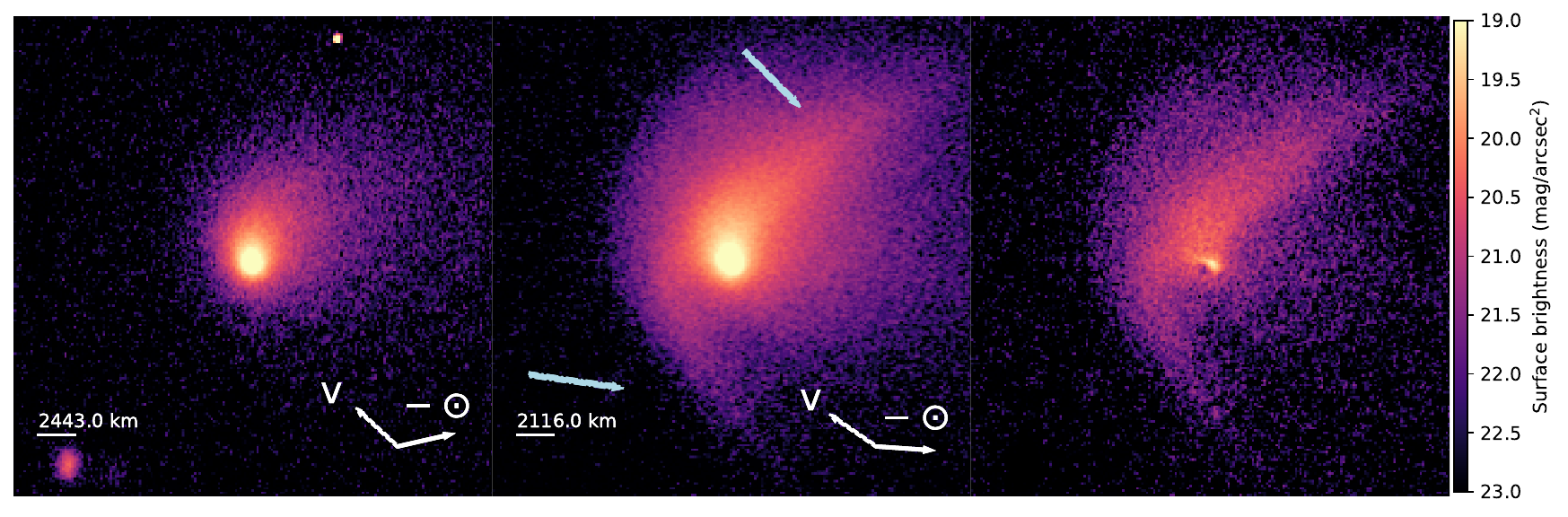}
        \includegraphics[scale=.50]{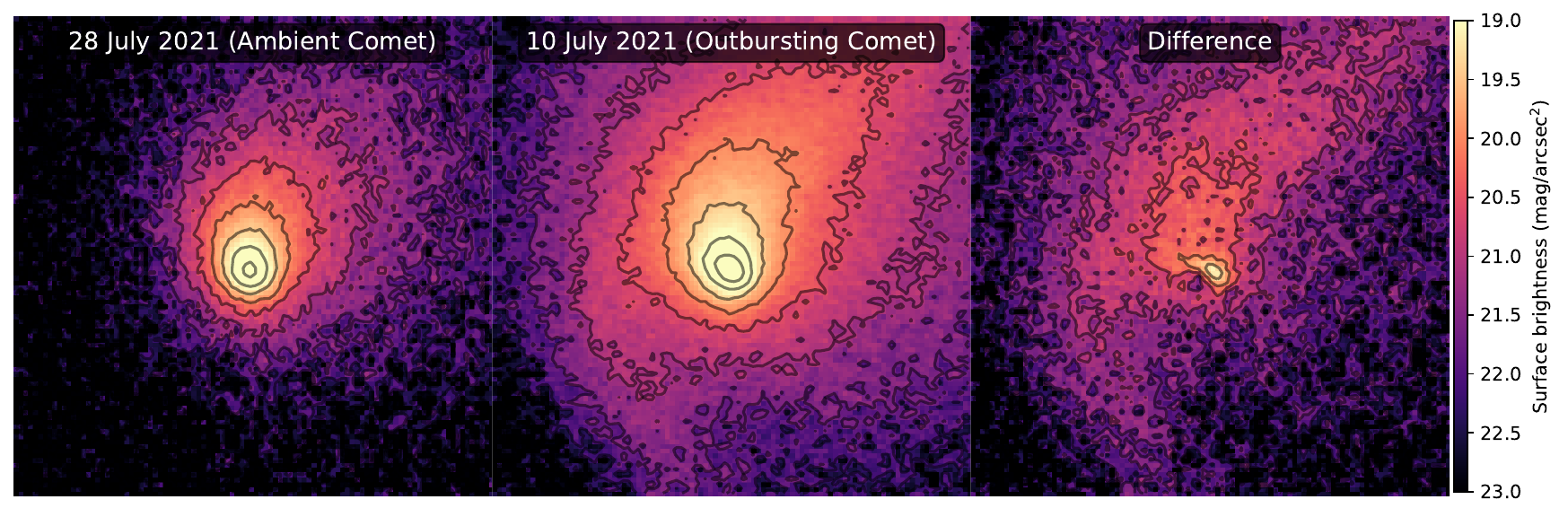}
        \caption{Images of comet 7P/Pons--Winnecke taken during a period of quiescence (left, taken at 2021 July 28 10:25:23 UTC) and the 2021 July 10 outburst (right, taken at 2021 July 10 02:29:58 UTC).  The outburst ejecta is apparent as distortions in the observed surface brightness as compared to the ambient coma image.  The bottom row shows the logarithmically scaled data with contours spaced every 0.5 mag/arcsec$^{2}$.  A color bar provides the surface brightness scale.  Images are oriented with north up, and east to the left. -$\odot$ represents the anti-solar direction and V is the heliocentric velocity direction, and a linear scale bar is shown. The blue arrows highlight two linear features as a result of the ejecta distribution that are meeting at a near-right angle in front of the nucleus.
        }
        \label{fig:jul10outburst}
    \end{figure}

\begin{figure}[ht]
        \centering
        \includegraphics[scale=0.50]{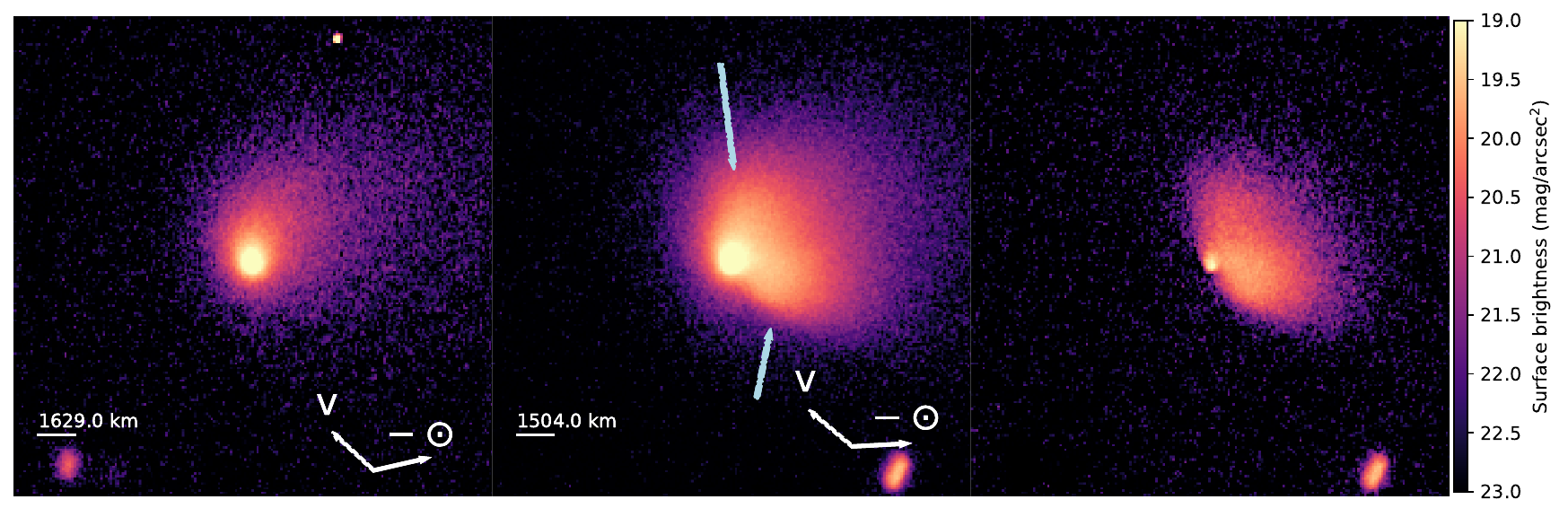}
        \includegraphics[scale=0.50]{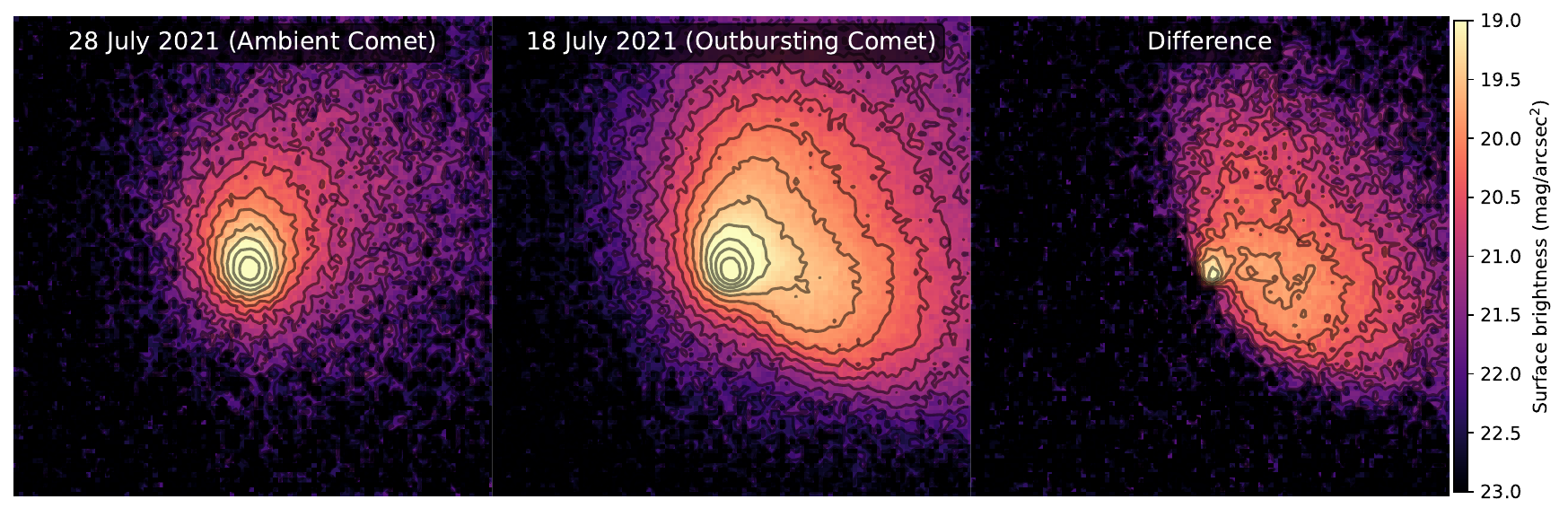}
        \caption{Same as Fig.~\ref{fig:jul10outburst}, but comparing the quiescent comet to the July 18 outburst (image taken on 2021 July 19 09:53:16 UTC, 13 hours after the comet was seen outbursting on July 18 (2021 July 18 22:17:16 UTC) where the comet's outbursting activity is more visible). Light blue arrows point at the two sides of an asymmetric cone of ejecta material with a broad opening angle of $\sim$130$^{\circ}$ pointing in the anti-sunward direction.}
        \label{fig:july18outburst}
    \end{figure}

    We compare the July 10 and 18 outbursts to an image of the comet in quiescence from July 28 in Figs.~\ref{fig:jul10outburst} and \ref{fig:july18outburst}.  The ambient coma has a prominent asymmetry centered on a position angle $\sim96$\degr{} north of the projected comet-Sun vector i.e., near-perpendicular to the sun.  The overall coma shows an anti-sunward (i.e., tail-ward) bias on $>$2000~km length scales.  The difference between the outburst image and the ambient image isolates the outburst material (Figs.~\ref{fig:jul10outburst} and \ref{fig:july18outburst}, right).

    The material ejected by the July 10 outburst is distributed over a broad range of position angles around the projected comet-Sun vector.  The ejecta has a peculiar distribution, with the appearance of two linear features meeting at a near-right angle.  Most of the ejecta is found to the north of the comet-Sun vector. The morphology of the ejecta enabled us to estimate its speed in the imaging plane.  We manually measured the motion of the corner and the southern line features (Fig.~\ref{fig:jul10outburst}) in unprocessed images taken on 2021 July 10 02:39 and 10:29 UTC.  Measured with respect to the nucleus, the corner moved from 6 to 23\arcsec{} and the line from 12 to 22\arcsec{} over the 7.8 hr period, corresponding to projected speeds of 220 $\pm$ 13~m~\inv{s} and 129 $\pm$ 13~m~\inv{s}, respectively.  A conservative 1\arcsec{} uncertainty (half the seeing) in the motion was assumed.  Based on the projected speeds, we measure the outburst onset time to be July 09 between 17:16 ($\pm$30~min) and 23:53 ($\pm$5~min).  The discrepancy between the two estimates could be due to differential projection effects, non-linear motion, or perhaps the outburst ejecta was produced over a $\sim$7~hr period.

    Comparing the July 10 and July 18 outbursts shows different dynamical behaviors. Assuming that the material in each event is initially the same and evolves in a similar manner, the differences are likely due to the orientation of their respective source locations.  Although in both events some material appears to be ejected towards the north, overall the July 18 material is distributed in the anti-sunward direction, whereas the July 10 event initially contained more sunward material.  We compared the sky coordinates of the comet in July 10 and 18 images to determine if the different ejecta distributions could be due to a change in projection effects. The comet only moved 5\degr{} on the sky, and a difference in the projection effects is unlikely to have a significant role.

    The age of the outbursts at the time of the images could explain the different morphologies. We define the outburst age of the comet as the time passed since the material was initially ejected. We examined the maximum outburst age at the time of the images presented in Figs.~\ref{fig:jul10outburst} and \ref{fig:july18outburst}.  The July 10 outburst was at most 44~hours old and the July 18 outburst at most 36 hours old.  Given that the general motion of the ejecta is initially radial and subsequently evolves tail-ward, it is unlikely that the younger July 18 event, with material already in the tail direction, could evolve into a phase similar to that of the older July 10 event, with material more distributed in the sunward hemisphere.

    With observational and temporal causes ruled out, we conclude that the primary difference between these two events is the orientation of the source location on the surface.  Different sources could be responsible, or the same source with the nucleus in a different rotational state.  Comet 7P has a measured period \citep[$7.9^{+1.6}_{-1.1}$~hr;][]{snodgrass05}. However, given the uncertainty in timing of the outburst onsets, we cannot determine if the same source region is producing both events.

\begin{figure}
        \centering
        \begin{interactive}{animation}{7p-look-movie-20210707T1536-20210723T1025.mp4}
            \includegraphics[width=\textwidth]{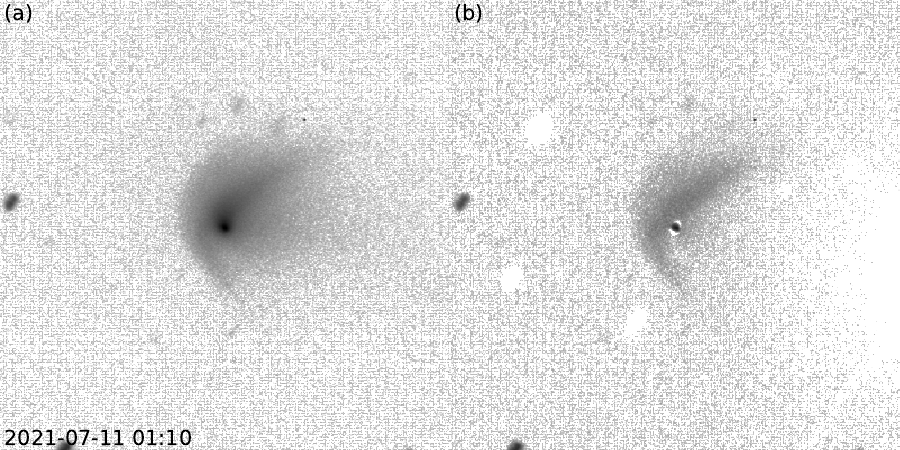}
        \end{interactive}
        \caption{Animation of comet 7P from 2021 July 07 to July 23 showing the July 10 and 18 outbursts.  (a) Original $r^\prime$ images.  (b) Temporally-filtered images generated by subtracting the data taken 2021 July 07 at 00:25 UTC.  Both panels are log-scaled from 24.2 to 18.0~mag/sq.~arcsec with a reversed gray-scale color map.  Orientation is the same as in Figs.~\ref{fig:jul10outburst} and \ref{fig:july18outburst}.  The example figure is the animation frame for 2021 July 07 at 01:10 UTC.}
        \label{fig:motion}
    \end{figure}

    Figure~\ref{fig:motion} is an animation of the data from 2021 July 07 to July 23, including the July 10 and July 18 outbursts, showing the original images and a temporally filtered image generated by subtracting a reference image for the ambient coma.  In both cases the ejecta establishes a characteristic morphology that fades with time.  For the July 10 event, the corner-shaped morphology grows in the sunward hemisphere until July 11 01:10 UTC.  The ejecta fades thereafter and appears to disperse towards the tail-ward hemisphere. The cone-shaped July 18 event, already directed in the tail-ward hemisphere, simply grows in extent while the surface brightness fades with time.

    \subsection{Ejecta mass}
    \label{sec:mass}
    The total mass $M$ of the outburst material is calculated from:
    \begin{equation}
        M = \int_{a_0}^{a_1} \frac{dn}{da} m(a) da
        \label{eqn:outburstmass}
    \end{equation}
    where $m(a)$ is the mass of a grain of radius $a$, $\frac{dn}{da}$ is the differential grain size distribution, and $a_0$, $a_1$ are the smallest and largest grain radii, assumed to be 0.1 and 1000~\micron, respectively.  We also assume spherical grains with a density of 1~g~\inv[3]{cm}.  The differential grain size distribution is taken to follow the form:
    \begin{equation}
        \frac{dn}{da} = N a^k
        \label{eqn:dn/da}
    \end{equation}
    where $k$ is the power-law slope ($-3.5$ is assumed), and $N$ is a normalization constant.  The normalization can be derived from the total geometric cross-sectional area, $G$, which is in turn based on the observed brightness
    \begin{equation}
        N = \frac{G}{\pi \int_{a_0}^{a_1}a^{(k+2)} da}
        \label{eqn:normconstant}
    \end{equation}
    \begin{equation}
        G = \frac{\pi r^2_h \Delta^2}{A_p \Phi(\phi)} 10^{-0.4(m - m_\odot)}
        \label{eqn:crosssecarea}
    \end{equation}
    where $m$ is the apparent magnitude of the dust ejecta, $m_\odot$=--26.93~mag is the apparent magnitude of the Sun at 1 au \citep{willmer18-sun}, and $A_p$ is the geometric albedo, assumed to be 4\%.  Here, the magnitudes and albedo are all evaluated for the $r$-band. Equation \ref{eqn:crosssecarea} ignores the dependence of scattering efficiency on grain size, but it is sufficient for an order of magnitude estimate.

    Several assumptions must be made to calculate the mass, which can have large consequences on the final estimate.  The differential size distribution power-law slope $k$ and the size range ($a_0$, $a_1$) dominate the uncertainty in the ejecta mass.  For example, values of $k>-3$ cause large particles to dominate the mass while the opposite is true for values $k<-4$.  Varying $k$ from $-3.0$ to $-4.0$ causes 7P's ejecta masses to change by a factor of 20  \citep[see also][]{tubiana15,gritsevich26-outbursts}. The ejecta masses, ranging from $1.2 \times 10^5$ kg to $2.1 \times 10^6$ kg with a median mass of $2.1 \times 10^5$ kg (Table~\ref{tab:outbursts}), are best used for comparison with other cometary outbursts and as an order of magnitude estimate that is based on several assumptions, rather than being taken as an absolute value with a proper uncertainty.  For consistency with previous works, our adopted size distribution parameters ($k=-3.5$, $a_0=0.1$~\micron, and $a_1=1$~mm) are the same as \citet{kelley21-wirtanen}, which produce masses in the middle of the two parameter sets of comet 67P from \citet{vincent16-fireworks}, and a factor of 2 lower than the dynamical model for comet 15P/Finlay by \citet{ishiguro16-outbursts}.  However, note that the parameters for comet 7P's quiescent coma by \citet{mariblanca-escalona25-7p} ($k=-3.7$, $a_0=5$~\micron, and $a_1=10$~cm) yield masses 23 times larger than our adopted values.  To emphasize the uncertainty in the mass estimate, we also quote results with these parameters in Table~\ref{tab:outbursts}.

    \section{Discussion}

    \subsection{Outburst frequency}
    \citet{kelley21-wirtanen} assumed the mini-outbursts (events ejecting $\lesssim10^6$~kg of dust) of comet 46P were related to terrain features and processes similar to those seen at comet 67P (Section~\ref{sec:intro}), and we proceed with the same assumption for 7P. It follows that the outburst frequency per nuclear surface area may correlate with the topographical relief of the comet as a whole \citep{kelley21-wirtanen}.

    \begin{deluxetable}{lcccccc}
        \tablecaption{Surface area normalized mini-outburst rates for comets.}
        \label{tab:frequency}
        \tablehead{
            \colhead{Comet}
            & \rh
            & \colhead{$N$}
            & \colhead{$t$}
            & \colhead{$f$}
            & \colhead{$A$}
            & \colhead{$f/A$}
            \\
            & (au)
            &
            & (day)
            & (\inv{day})
            & (km$^2$)
            & (\inv{day}\inv[2]{km})
        }
        \startdata
        67P/Churyumov--Gerasimenko\tablenotemark{a} & 1.2--1.4 & 34 & 91 & 0.37 & 46.9 & 0.008 \\
        46P/Wirtanen & 1.1--2.3 & 6 & 388 & 0.015 & 3.94 & 0.004 \\
        103P/Hartley 2\tablenotemark{a} & 1.1 & 0 & 50 & 5.24 & $<$0.02 & $<$0.004 \\
        9P/Tempel 1\tablenotemark{a} & 1.5 & 10 & 33 & 0.30 & 108 & 0.003 \\
        41P/Tuttle--Giacobini--Kres\'ak & 1.0--1.6 & 1 & 128 & 0.008 & $4.5\pm0.6$ & 0.002  \\
        7P/Pons--Winnecke & 1.2--1.7 & 10$\pm$1 & 89 & 0.011 & $85\pm13$ & 0.001 \\
        67P/Churyumov--Gerasimenko (2022)\tablenotemark{b} & 1.2 & 2 & 65 & 0.03 & 46.9 & 0.0007 \\
        67P/Churyumov--Gerasimenko (2022)\tablenotemark{b+c} & 1.2--2.6 & 2 & 269 & 0.007 & 46.9 & 0.0002 \\
        49P/Arend--Rigaux & 1.8--2.5 & 1 & 33 & 0.030 & 289 & 0.0001 \\
        67P/Churyumov--Gerasimenko (2015)\tablenotemark{c} & 1.2--3.3 & 1 & 437 & 0.002 & 46.9 & $5\times10^{-5}$ \\
        \enddata
        \tablecomments{$N$ is number of mini-outbursts, $t$ is the time period analyzed for outbursts, $f$ is frequency, $A$ is nuclear surface area.  9P, 67P, and 103P were previously calculated by \citet{kelley21-wirtanen}.  Outburst and surface area references: 7P, \citet{fernandez13}; 9P, \citet{belton08}, \citet{thomas13-tempel1}; 41P, \citet{boehnhardt20-tgk}; 49P, \citet{millis88}, \citet{eisner17-49p}; 67P, \citet{boehnhardt02-wirtanen}, \citet{kelley21-wirtanen}; 67P, \citet{vincent16-fireworks}, \citet{jorda16}; 103P, \citet{ahearn11}, \citet{meech11-epoxi}, \citet{thomas13-hartley2}.}
        \tablenotetext{a}{Based on high-cadence spacecraft observations.  Other rows are from ground-based data.}
        \tablenotetext{b}{\citet{sharma21-67p}}
        \tablenotetext{c}{\citet{gardener22}}
    \end{deluxetable}

    Based on the estimated size of 7P \citep[$R=2.6\pm0.2$~km;][]{fernandez13} the surface area of an equivalent sphere is $85\pm13$~km$^2$, which yields an area-normalized outburst rate of 0.001~\inv{day}~\inv[2]{km} during our 89~day period of coverage. Comets 9P, 46P, 67P, and 103P were previously studied by \citet{kelley21-wirtanen}, and we present those results along with 7P in Table~\ref{tab:frequency}.  We also add calculations for comets 41P/Tuttle-Giacobini-Kres\'ak \citep{boehnhardt20-tgk}, and 49P/Arend-Rigaux \citep{eisner17-49p}.  Most comets, including comet 7P, have mini-outburst rates of order $10^{-3}$~\inv{day}~\inv[2]{km}.  If they are caused by topographical failure (e.g., cliffs), then this suggests their surfaces are topographically similar to the average surface of comet 9P, as opposed to the high outburst rate and frequent steep terrain features of comet 67P \citep{vincent17-evolution}.  However, if they are instead driven by the explosive outgassing of subsurface \coo{} reservoirs explosions, then the interpretations are not as clear.  The correlation between 67P's mini-outbursts and topography has been established, but at this time we cannot preclude \coo{} driven outbursts in smooth terrain on other comets.  More work is needed to understand how \coo{} subsurface cavities form \citep{mueller25-outbursts}.

    There is a caveat to our analysis of Table~\ref{tab:frequency}: that 67P, 9P, and 103P were all observed by intense high-cadence observing sequences from spacecraft.  It is possible, that the measured outbursts rates of these comets are enhanced compared to what we can observe from the ground.  To consider this further, we examine the results of \citet{gardener22}, who produced a lightcurve of 67P from 26 different sources, uniformly processed and analyzed.  With photometric data covering a period of about 450 days in 2015/2016, they only found 1 outburst with a relative strength of --0.14~mag, and a mass estimate of $\sim2\times10^5$~kg.  We also add to Table~\ref{tab:frequency} calculations for observations taken during the more favorable 2021 perihelion passage from \citet{gardener22} and \citet{sharma21-67p}.  The ground-based results have outburst rates $\sim10^{-5}$ to $\sim10^{-4}$~\inv{day}~\inv[2]{km}.  In agreement with the discussion of \citet{gardener22}, these results call into question whether the high outburst rate of 67P as seen by Rosetta is a fair benchmark for the ground-based data.

    \subsection{Comparison to meteor outburst constraints}

    Meteor activity provides a record of past cometary dust production (see \citealt{jenniskens06}, \citealt{trigo-rodriguez22}, \citealt{ye24-comets3}, and references therein). By combining meteor observations with dynamical models, we can constrain models of the past activity of a comet. The June Bo\"otids, the meteor shower associated with 7P, is largely inactive in most years but has exhibited significant activities in 1916, 1998 and 2004, due to Earth encountering cometary dust ejected in the past \citep{ye24-comets3}. Previous dynamical models have shown that these meteor outbursts were produced by cometary dust ejected between 1813 and 1875 \citep[cf.][]{vaubaillon05-bootids}. The observed meteoroid flux in 1998 corresponds to a mass loss of $6\times10^8$~kg over one orbit \citep{ye24-comets3}.

    Is it possible to use the observed meteor outbursts to constrain 7P's outbursts in the 1800s? To explore this question, we simulate the formation and evolution of the June Bo\"otid meteoroid stream using the approach described in \citet{ye16-209p}. In brief, we take the orbital elements of 7P in 1819 (JPL orbit SAO/1819), integrate it back to 1801, and then integrate forward again, releasing meteoroids when the comet is within 2~au from the Sun (consistent with observations of the comet by \citealt{kelley21-atel14486} and \citealt{novichonok24-7p}). We use the 1819 orbit rather than integrating from the current orbit, as the former more closely represents 7P's orbit in the 1800s when the meteor outbursts' materials were produced. We limit the simulation back to 1801 because 7P had a close encounter with Jupiter on 1800 Aug 27 (by 0.11~au), significantly altering its orbit and rendering earlier orbits very uncertain.

    Meteoroids are released in the sunlit direction, following the water ice sublimation model described by \citet{whipple50-encke}, and are integrated along with the comet using a Gragg-Bulirsch-Stoer integrator implemented in a modified Mercury6 program \citep{chambers99, ye16-209p}. The integration includes the Sun, the eight major planets with the Earth-Moon system represented by a single particle at their barycenter, and radiative effects on the meteoroids (radiation pressure and Poynting-Robertson drag). We use the contemporary measured $Af\rho=149 r_\mathrm{h}^{-0.4}$~cm \citep{lister22-look} to estimate the meteoroid flux by relating \afr{} to a mass-loss rate for a given dust size distribution, following the methods described by \citet{ye14-209p-predictions} and \citet[][\S~3.2]{ye16-209p}. This assumes symmetric dust production with respect to heliocentric distance. Although the long-term light-curve showed an asymmetric dust production (compare \citealt{lister22-look} to \citealt{kelley21-atel14486}), our cursory tests showed that this difference would not affect the results in significant levels: the timing and meteoroid flux would be within a fraction of an hour and 10--20\%, which are well within the FWHMs and typical error of meteor outburst forecasts \citep{egal20-forecasting}. The simulation models meteoroids in the 0.5--5~cm diameter range, assuming a bulk density of $500~\mathrm{kg~m^{-3}}$, which corresponds to the size regime producing visual meteors. Simulated dust and meteoroids follow a size distribution of $N(a) \propto a^{-3.5}$, consistent with measurements from the 1998 June Bo\"otid outburst \citep{arlt99-bootids98} and the dust tail analysis of \citet{mariblanca-escalona25-7p}.

    Comparing the simulation results with the observed meteor outbursts (Table~\ref{tbl:met-obs}), we find several notable outcomes:

    \begin{itemize}
        \item The well-observed meteor outbursts in 1916, 1998 (the ``late'' component peaked near Jun 27 at 21h UT), and 2004 are well-reproduced by our simulation. The calculated timing of the outbursts are within the FWHM of the observed activity. The calculated Zenith Hourly Rate (ZHR; a proxy of meteoroid flux) agrees with the observed values within a factor of 2, which is reasonable compared to other well-modeled meteor outbursts \citep[cf.][]{egal20-forecasting}.
        \item While there are slight divergences in details, our simulation generally agrees with earlier work \citep{vaubaillon05-bootids} that these meteor outbursts were produced by 7P's activity in the early 1800s. The differences between our model and earlier models likely arise from using different initial orbits for the simulations: the three models presented in \citet{vaubaillon05-bootids} used orbits from 7P's 1996 and 2002 returns, integrating them backward to 1802, whereas we used orbits derived from data between 1819 and 1858, resulting in small but noticeable differences in 7P's orbits in the 1800s.
        \item The poorly observed ``early'' peak of the 1998 outburst (occurring 8 hours before the late peak) remains unaccounted for. Additionally, we did not reproduce the suspected meteor activities observed in 1921 and 1927 \citep[cf.][]{arlt99-bootids98}.
        \item Our simulation also suggests significant activities in 1910, 1941, 1960--61, and 1986, none of which were reportedly detected.
    \end{itemize}

    \begin{deluxetable*}{lcccclccl}
        \tablecaption{Significant (ZHR$>10$) modeled June Bootid outbursts and the observed activities since 1900.}
        \label{tbl:met-obs}
        \tablecolumns{8}
        \tablehead{
            \multicolumn{4}{c}{Modeled} & & \multicolumn{4}{c}{Observed} \\
            \cline{1-4} \cline{6-9}
            \colhead{Peak time (UT)} & \colhead{FWHM} & \colhead{Ejection year} & \colhead{$ZHR$} & & \colhead{Peak time (UT)} & \colhead{FWHM} & \colhead{$ZHR$} & \colhead{Source}
        }
        \startdata
        1910 Jun 29 18:09 & 3~hr & 1813 & 90 & & \multicolumn{4}{c}{(Not reported)} \\
        1916 Jun 28 23:17 & 3~hr & 1813, 1819 & 150 & & 1916 Jun 28 22:30 -- Jun 29 01:00 & $\gtrsim3$~hr & $\sim100$ & A99 \\
        1916 Jul 3 23:08 & 4~hr & 1847 & 40 & & \multicolumn{4}{c}{(Not reported)} \\
        \multicolumn{4}{c}{(No encounter)} & & 1921 Jun 29 & n/a & $\leq10$ & A99 \\
        \multicolumn{4}{c}{(No encounter)} & & 1927 Jun 24--26, Jun 29 to Jul 2 & n/a & $\leq30$ & A99 \\
        1941 Jul 6 16:43 & 5~hr & 1819, 1825 & 90 & & \multicolumn{4}{c}{(Not reported)} \\
        1960 Jul 5 23:18 & 4~hr & 1825, 1836 & 60 & & \multicolumn{4}{c}{(Not reported)} \\
        1961 Jul 6 07:07 & 5~hr & 1825, 1836 & 40 & & \multicolumn{4}{c}{(Not reported)} \\
        1986 Jul 1 09:36 & 6~hr & 1830--1841 & 50 & & \multicolumn{4}{c}{(Not reported)} \\
        (1998 Jun 27 14:03) & (2~hr) & (1847$^\dag$) & (10) & & 1998 Jun 27 12:30 & 4~hr & $250\pm40$ & HO98, A99b \\
        1998 Jun 27 19:39 & 8~hr & 1808--1847 & 180 & & 1998 Jun 27 21:00 & 3~hr & $81\pm7$ & A99 \\
        2004 Jun 23 14:01 & 4~hr & 1813--1847 & 60 & & 2004 Jun 23 14:46 & 6~hr & $33\pm10$ & V05 \\
        \enddata
        \tablecomments{References: A99 - \citet{arlt99-bootids98}; A99b - \citet{arlt99-bootids}; HO98 - \citet{hashimoto1998june}; V05 - \citet{vaubaillon05-bootids}. \dag: association uncertain (see main text); observed properties based on data from a single observer.}
    \end{deluxetable*}

    The source of the 1916 meteor outburst is relatively simple and consistent between models: the outburst material is predominantly composed of 7P's ejecta from 1819 (with $\sim85\%$ contribution). Our model indicates that the meteor outburst material was produced throughout 7P's active period during its 1819 orbit; planetary dynamics did not preferentially deliver material from a specific part of that orbit to Earth. Cursory checks suggest this is also true for the contributing ejecta to the 1998 and 2004 meteor outbursts. Assuming an outburst FWHM of 2~hrs and a representative meteoroid diameter of 1~cm (corresponding to $V\sim3$~meteors), we derive a dust production of $2\times10^8$~kg for that orbit, comparable to the value found in \citet{ye24-comets3} for the 1825 ejecta. Similar to the discussion in Section~\ref{sec:mass}, this number heavily depends on the assumptions made and should be considered as an order of magnitude approximation.

    We conclude that back in the early 1800s, 7P did not experience outbursts significantly larger ($\gtrsim10^8$~kg level) than those reported in this study, as such events would have produced stronger meteor activities than observed. This conclusion aligns with our earlier finding that 7P's outbursts are mostly small ($\lesssim10^7$~kg).

    The relatively accurate reproduction of meteoroid flux in our model is notable and suggests that 7P's activity level has not changed dramatically from the early 1800s to the present, even though its orbit has evolved from $q=0.77$~au in 1819 to $q=1.24$~au in 2015.

    However, our model remains unable to explain certain observed meteor outbursts, most notably the earlier peak of the 1998 outburst. Our model shows that Earth passed the 1847 ejecta at 14:03 UT of 1998 Jun 27, suggesting that this event could theoretically be explained by a large outburst of 7P during its 1847 apparition that was $25\times$ larger than the baseline activity, corresponding to $\sim10^9$-kg level. However, this scenario would also produce very strong meteor activity (with $ZHR\sim1000$) on 1916 Jul 3, when Earth passed the same ejecta at that time. This event was well-placed for observation from Europe but was not reportedly detected. While it is possible that the outburst occurred but was missed, a more likely scenario is that the earlier outburst in 1998 might have been produced by an older ejecta not included in our model. We defer further investigation of this question to future research, as exploring 7P's uncertain orbit before 1800 is beyond the scope of this paper.

    \section{Conclusions and Summary}

    We studied eight outbursts of comet 7P/Pons--Winnecke in a high-cadence dedicated outburst discovery campaign between 2021 June 3 and 2021 August 31 and computed an outburst rate of 0.09 to 0.13~\inv{day} (98\% confidence interval) for the same period.  The outburst ejecta had apparent brightnesses ranging from 16.2 to 13.4~mag. The nucleus surface area normalized outburst rate during this time period was 0.001~\inv{day}~\inv[2]{km}.

    The interpretation of the outburst rate depends on the causes of the events.  Recent work suggests there are at least two mini-outburst drivers at 67P: terrain failure and exposure of water ice versus the explosive release of \coo{} \citep{mueller24-outbursts}.  But, the interpretation of the outburst rates in other comets is uncertain.  The outburst rate of 7P, being 8 times lower than 67P, could suggest that its terrain is topographically eroded, perhaps more like 103P/Hartley~2 than 67P/Churyumov-Gerasimenko.  Alternatively, it may be related to each comet's \coo{} abundance or the depth at which \coo{} is buried and the thermophysical properties of the nucleus subsurface.  More work on the causes of the \coo{}-driven outbursts is needed \citep{mueller25-outbursts}.

    The morphologies of the outburst ejecta varied, with no apparent preference for direction on the sky, including the projected solar direction. The most likely explanation is that the events are from distinct sources, perhaps distributed across the surface of the nucleus. An analysis of the motion of the ejecta from one of the larger outbursts (2021 July 10) yielded projected speeds of $\sim100-200$~m~\inv{s}, and the other outbursts had lower limit speeds of 20--70~\mps. An order of magnitude calculation of the ejecta masses for each outburst yields a range from $\sim10^5$ to $10^6$~kg. These mass estimates are smaller than the $\sim10^7$~kg ejected by the Deep Impact mission to comet 9P/Tempel~1 \citep{ahearn05,ahearn08}, and similar to the mini-outbursts of comet 9P, 46P/Wirtanen, and 67P/Churyumov-Gerasimenko \citep{belton08,vincent16-fireworks,kelley21-wirtanen}.

    Observations and modeling of the associated June Bo\"otid meteor shower suggests that some meteor shower outbursts can be explained with quiescent activity at 7P's current levels.  As a result, 7P likely did not produce cometary outbursts with $\sim10^8$~kg in the early 1800s (about 30--40 orbits ago).  However, some meteor shower outbursts could not be explained by our model.  We suggest they may be produced by activity before the 1801 orbital perturbation by Jupiter.

    Further studies may provide more insight on mini-outbursts and outbursts in general.  The order of magnitude discrepancy between the Rosetta-observed outburst rate of 67P and the ground-based observed rate remains unresolved.  Regardless, there seems to be a range of outburst rates in the short-period comet population, which may be a consequence of surface topography and (erosional) age \citep{guilbert-lepoutre23-pits}.   Although similar observations of 7P may not be repeated for some time (the next close approach to Earth will not be until 2045)\footnote{NASA JPL Small-Body Database: \url{https://ssd.jpl.nasa.gov/tools/sbdb_lookup.html\#/?sstr=7P}}, high-cadence observations of other comets will be useful in comparison, as will observations of future spacecraft targets.  Time-domain surveys, such as Vera C. Rubin Observatory's Legacy Survey of Space and Time (LSST) \citep{ivezic19-lsst,bianco22-optimization}, will provide the data needed to discover new outbursts \citep{schwamb23-cadence} and focus follow-up resources so that even higher cadence observations appropriate for mini-outburst discoveries and spectroscopic follow-up will be fruitful.

    \begin{acknowledgements}
        The authors thank M. Knight for reviewing an early version of the manuscript and providing helpful comments.

        M.S.P.K. acknowledges funding from NASA (USA) Solar System Observations program grant 80NSSC20K0673.  M.E.S acknowledges support from UK Science and Technology Facilities Council (STFC) grant ST/X001253/1.

        This work makes use of observations from the Las Cumbres Observatory (LCO) global telescope network.

        Some observations were made by members of the Comet Chasers education and outreach project, which is part of the DeepSpace2DeepImpact project funded by STFC(UK).  Access to the LCO telescopes was provided by the Faulkes Telescope Project (supported by the Dill Faulkes Educational Trust) (FTPEPO2014A-004).  The Comet Chasers observers included Tony Angel, Wim Cuppens, Richard Miles, and Helen Usher.

        This paper is based on observations made with the MuSCAT3 instrument, developed by the Astrobiology Center and under financial supports by JSPS KAKENHI (JP18H05439) and JST PRESTO (JPMJPR1775), at Faulkes Telescope North on Maui, HI, operated by the Las Cumbres Observatory.

        The Digitized Sky Survey was produced at the Space Telescope Science Institute under U.S. Government grant NAG W--2166. The images of these surveys are based on photographic data obtained using the Oschin Schmidt Telescope on Palomar Mountain and the UK Schmidt Telescope. The plates were processed into the present compressed digital form with the permission of these institutions.

        This research made use of the Horizons online ephemeris system developed and operated by the Solar System Dynamics Group at NASA's Jet Propulsion Laboratory.
    \end{acknowledgements}

    \vspace{5mm}
    \facilities{LCOGT, FTN, MAST (ATLAS Refcat2, DSS)}

    \software{astropy \citep{astropy13, astropy18, astropy22},
        sbpy \citep{mommert19-sbpy},
        photutils \citep{bradley22-photutils},
        matplotlib \citep{hunter07-mpl}
    }

    \bibliography{apj-jour,references,mikes-reference-file}
    \bibliographystyle{aasjournal}

\end{CJK*}
\end{document}